\documentclass[12pt]{article}

\usepackage[reqno]{amsmath}
\usepackage{mathrsfs}
\usepackage{amssymb}
\usepackage{verbatim}
\usepackage{rotating} 

\usepackage{bbm}
\usepackage{array}
\usepackage{float}
\usepackage{color}
\usepackage{graphicx}
\usepackage{a4}
\usepackage{a4wide}
\usepackage{wasysym}

\newcommand{\dms}{\mbox{$\Delta m^2_{21}$}}
\newcommand{\dma}{\mbox{$|\Delta m^2_{31}|$}}

\newcommand{\onbb}{\mbox{$0\nu\beta\beta$}}

\newcommand{\be}{\begin{eqnarray}}
\newcommand{\ee}{\end{eqnarray}}
\newcommand{\beq}{\begin{equation}}
\newcommand{\eeq}{\end{equation}}

\newcommand{\eps}{\mbox{$\epsilon$}}

\begin{document}

\setlength{\unitlength}{1mm}

\begin{titlepage}
\title{\vspace*{-2.0cm}
\bf\Large
Constraining New Physics with a Positive or Negative Signal of Neutrino-less Double Beta Decay
\\[5mm]\ }

\author{
Johannes Bergstr\"om\thanks{email: \tt johbergs@kth.se}~, Alexander Merle\thanks{email: \tt amerle@kth.se}~, and Tommy Ohlsson\thanks{email: \tt tommy@theophys.kth.se}
\\ \\
{\normalsize \it Department of Theoretical Physics, School of Engineering Sciences,}\\
{\normalsize \it Royal Institute of Technology (KTH),}\\
{\normalsize \it Roslagstullsbacken 21, 106 91 Stockholm, Sweden}
}
\date{\today}
\maketitle
\thispagestyle{empty}

\begin{abstract}
\noindent
We investigate numerically how accurately one could constrain the strengths of different short-range contributions to neutrino-less double beta decay in effective field theory. Depending on the outcome of near-future experiments yielding information on the neutrino masses, the corresponding bounds or estimates can be stronger or weaker. A particularly interesting case, resulting in strong bounds, would be a positive signal of neutrino-less double beta decay that is consistent with complementary information from neutrino oscillation experiments, kinematical determinations of the neutrino mass, and measurements of the sum of light neutrino masses from cosmological observations. The keys to more robust bounds are improvements of the knowledge of the nuclear physics involved and a better experimental accuracy.
\end{abstract}

\end{titlepage}

\section{\label{sec:Intro}Introduction}

Within the last decades, it has become more and more apparent that neutrinos can be used as one of the major windows to look for physics beyond the Standard Model (SM) of elementary particles, which is often dubbed ``new physics''. Not only has it been established by neutrino oscillation experiments on solar~\cite{Abdurashitov:2002nt,Altmann:2005ix,Aharmim:2007nv,Arpesella:2007xf}, atmospheric~\cite{Ashie:2005ik}, and artificially produced neutrinos~\cite{KamLAND:2008ee,Ahn:2006zza,Adamson:2007gu} that neutrinos are massive, but also the corresponding leptonic mixing differs significantly from the quark sector~\cite{Nakamura:2010zzi}. Apart from their values, one of the major questions about neutrino masses is their origin. A particularly attractive possibility to explain their smallness is the so-called ``seesaw mechanism''~\cite{Minkowski:1977sc,Yanagida:1979as,Glashow:1979nm,GellMann:1980vs,Mohapatra:1979ia}. This mechanism, however, intrinsically involves the breaking of lepton number by two units. If this breaking was indeed realized in Nature, a process that would have to appear is neutrino-less double beta decay ($0\nu\beta\beta$), where an even-even nucleus undergoes a transition $(A,Z)\to (A,Z+2) + 2 e^-$, thereby violating lepton number. This process has been intensively searched for, but up to now there has not yet been an unambiguous detection~\cite{Aalseth:2004hb} (cf., however, the claim by part of the Heidelberg-Moscow collaboration in Ref.~\cite{KlapdorKleingrothaus:2004wj}).

Although the ``standard'' mechanism for \onbb\ is the exchange of light Majorana neutrinos~\cite{Doi:1985dx,Bilenky:1987ty,Schmitz:1997ag}, other mechanisms could very well appear in certain models beyond the SM, such as supersymmetry (see, e.g., Refs.~\cite{Mohapatra:1986su,Allanach:2009iv}) or left-right symmetric models (see, e.g., Refs.~\cite{Mohapatra:1986pj,Tello:2010am}). To disentangle the various possibilities, it will probably be necessary to detect the process for several different isotopes~\cite{Deppisch:2006hb,Gehman:2007qg}.

In this paper, we will investigate point-like contributions of new physics to \onbb, previously studied in Refs.~\cite{Pas:2000vn,Prezeau:2003xn}. Over the last years, there has been considerable improvement in the knowledge of the neutrino oscillation parameters~\cite{Schwetz:2011qt}, and new experiments on neutrino masses and mixings are currently starting or will start in the near future. One of the most prominent near future experiments on neutrino-less double beta decay is certainly the GERDA experiment~\cite{Abt:2004yk}, which we will take as an example. This experiment will be complemented by others aiming to detect the kinematical electron-neutrino effective mass $m_\beta$ (e.g.\ KATRIN~\cite{Osipowicz:2001sq}), the cosmological sum $\Sigma$ of neutrino masses (e.g.\ Planck~\cite{Planck:2011ah}), or the leptonic mixing angle $\theta_{13}$ (e.g.\ Double Chooz~\cite{Ardellier:2004ui}). We therefore feel that it is necessary to investigate the impact of these experiments on new physics contributions to \onbb. Naturally, once this process is observed, there will soon be analyses of the standard picture. Later on, however, the question will arise what else we could learn from a positive or negative signal of \onbb. Here, we are going to give an answer to this question in what regards the short-range contributions to the process.

The paper is organized as follows: In Secs.~\ref{sec:EFT} and~\ref{sec:chi2}, we will shortly review the nature of the short-range contributions to \onbb, as well as the analysis of a signal under the assumption of the standard mechanism of light neutrino exchange. Our actual investigations will be presented in Secs.~\ref{sec:scenarios} and~\ref{sec:Rates}, where we first define certain benchmark scenarios and use them later on to illustrate how strong the bounds on the effective operators of the new physics contributions to \onbb\ could be for certain realistic situations after the future experiments have yielded results. Depending on if the signal of \onbb\ is positive or negative, and also depending on if it is consistent or inconsistent with results from complementary experiments, the bounds can be stronger or weaker. In general, it is hard to derive strict statistical bounds, due to the intrinsic difficulties arising from the nuclear physics involved. Finally, we present our conclusions in Sec.~\ref{sec:Conc}.

\section{\label{sec:EFT}Point-like contributions to $\boldsymbol{0\nu\beta\beta}$}

The key point of our analysis is that one can parametrize many different contributions of new physics to $0\nu\beta\beta$ in a nearly model-independent way by using effective field theory~\cite{Georgi:1994qn,Pich:1998xt,Manohar:1996cq}. This is a very powerful approach as long as the detailed spectrum of new heavy particles does not play a significant role.\footnote{See, e.g., Ref.~\cite{Blennow:2010th} for a case where the detailed heavy spectrum {\it does} play a role.} Depending on if one only modifies the vertices that couple to the light neutrinos or if the whole decay mechanism is altered, one obtains long-range~\cite{Pas:1997fx} and short-range~\cite{Pas:2000vn} contributions, respectively. Here, we restrict ourselves to the latter ones, since modifications of the vertices only are even discussed in some textbooks, see e.g.\ Ref.~\cite{Schmitz:1997ag}, and furthermore such modified weak interaction charged current vertices can be probed much more easily in, e.g., neutron beta decay experiments~\cite{Severijns:2006dr} or in muon conversion experiments~\cite{LeeRoberts:2007gf}.

When looking at the short range contributions, one has to study the following general Lagrangian~\cite{Pas:2000vn}:
\begin{equation}
 \mathcal{L}=\frac{G_F^2}{2}m_p^{-1} \left( \epsilon_1 JJj + \epsilon_2 J^{\mu\nu} J_{\mu\nu} j + \epsilon_3 J^\mu J_\mu j + \epsilon_4 J^\mu J_{\mu\nu} j^\nu + \epsilon_5 J^\mu J j_\mu \right) + \text{h.c.}
 \label{eq:L_0nbb}
\end{equation}
Here, $G_F$ is the Fermi constant, which is a reflection of the double weak interaction vertex involved in the standard light neutrino exchange diagram for \onbb, and $m_p$ is the proton mass, which is used to give the Lagrangian the correct mass dimension. In principle, any high energy scale could have served this purpose, but a scale of the order of the proton mass is already considerably higher than the typical internal momentum in the nucleus, which is of the order of 100~MeV~\cite{Bilenky:1987ty}. The strengths of the different operators are parametrized by the (generally complex) dimensionless coefficients $\epsilon_i$. The hadronic currents in Eq.~\eqref{eq:L_0nbb} are given by
\begin{equation}
 J_{L,R} = \overline{u}\left( 1 \mp \gamma_5 \right) d, \quad J^\mu_{L,R} = \overline{u} \gamma^\mu \left( 1 \mp \gamma_5 \right) d, \quad  J^{\mu\nu}_{L,R} = \overline{u} \frac{i}{2} \left[\gamma^\mu, \gamma^\nu\right] \left( 1 \mp \gamma_5 \right) d,
 \label{eq:hadron_operators}
\end{equation}
and the leptonic ones are
\begin{equation}
 j_{L,R} = \overline{e}\left( 1 \mp \gamma_5 \right) e^c = 2\ \overline{e_{R,L}} (e_{R,L})^c, \quad j^\mu_{L,R} = \overline{e} \gamma^\mu \left( 1 \mp \gamma_5 \right) e^c = 2\ \overline{e_{L,R}} \gamma^\mu (e_{R,L})^c.
 \label{eq:electron_operators}
\end{equation}
For all currents, different chirality structures are permitted. Note that we have also assigned the correct chirality to the electron operators in Eq.~\eqref{eq:electron_operators}, since we will need this for our argumentation later on.

Actually, we could have added some more Lorentz-invariant terms to Eq.~\eqref{eq:L_0nbb},
\begin{equation}
\mathcal{L}'=\frac{G_F^2}{2}m_p^{-1} \left(  \epsilon_6 J^\mu J^\nu j_{\mu\nu} + \epsilon_7 J J^{\mu\nu} j_{\mu\nu} + \epsilon_8 J_{\mu\alpha} J^{\nu\alpha} j_\nu^\mu \right),
 \label{eq:Lprime_0nbb}
\end{equation}
where the leptonic tensor current is given by $j^{\mu\nu}_{L,R} = \overline{e} \frac{i}{2} \left[\gamma^\mu, \gamma^\nu\right] \left( 1 \mp \gamma_5 \right) e^c$. These terms had been included in the Lagrangian in Ref.~\cite{Pas:2000vn}, but were neglected in the final analysis, since these authors worked in the $s$-wave approximation where such contributions vanish. In Ref.~\cite{Prezeau:2003xn}, however, it was pointed out that all operators proportional to $\overline{e}\gamma_\mu e^c$, $\overline{e}\frac{i}{2}\left[\gamma_\mu,\gamma_\nu\right] e^c$, and $\overline{e}\gamma_5 \frac{i}{2}\left[\gamma_\mu,\gamma_\nu\right] e^c$ vanish identically, since the electron fields are Grassmann numbers. Therefore, the terms in Eq.~\eqref{eq:Lprime_0nbb} are not relevant for \onbb.

The purpose of the effective field theory approach is to provide an easy and unified description of new physics. In particular, since there is only a finite set of operators, it allows to efficiently compare different underlying models in what regards their interaction strengths. Very recent works covering different explicit $0\nu\beta\beta$ mechanisms and in particular their interplay with the standard light neutrino exchange~\cite{Faessler:2011qw,Faessler:2011rv} appeared nearly simultaneously to this paper. All the mechanisms treated there can be translated into our effective language.

To give one explicit and illustrative example, we take the gluino exchange mechanism in $R$-parity violating supersymmetry (RPV-SUSY)~\cite{Faessler:2011qw,Faessler:2011rv,Chemtob:2004xr}. The goal is to reproduce Eq.~\eqref{eq:L_0nbb} by integrating out all heavy particles, and it is easy to see that the corresponding effective Lagrangian is given by
\begin{equation}
 \mathcal{L}_{\rm RPV}=\frac{G_F^2}{2}m_p^{-1} (\epsilon^{RRR}_1)_{\rm RPV}\ J_R J_R j_R + \text{h.c.},
 \label{eq:L_RPV}
\end{equation}
whose coefficient can be expressed in terms of model parameters as~\cite{Faessler:2011qw}
\begin{equation}
 (\epsilon^{RRR}_1)_{\rm RPV} = \frac{\pi \alpha_s}{6} \frac{(\lambda'_{111})^2}{G_F^2 m_{\tilde d_R}^4} \frac{m_p}{m_{\tilde g}} \left[ 1+ \left( \frac{m_{\tilde d_R}}{m_{\tilde u_L}} \right)^2 \right]^2,
 \label{eq:eps_RPV}
\end{equation}
where $\alpha_s=g_3^2/(4\pi)$ is the strong fine structure constant, $m_{\tilde u_L,\tilde d_R,\tilde g}$ are the masses of the up-squark, of the down-squark, and of the gluino, respectively, and the parameter $\lambda'_{111}$ encodes the strength of the RPV-SUSY operators. Obviously, if we have any kind of bound on the general coefficient $\epsilon^{RRR}_1$, this can be translated into a bound on the parameter $\lambda'_{111}$, and hence on the size of the new physics contribution.

Later on, we will derive bounds on $\lambda'_{111}$ for all scenarios under consideration, in order to illustrate how one can translate our results into bounds on concrete models of new physics. In order to arrive at numerical values for $\lambda'_{111}$, we will assume that $m_{\tilde u_L}=m_{\tilde d_R}=800$~GeV and $m_{\tilde g}=1000$~GeV. These values are good examples for the Minimal Supersymmetric Standard Model squark and gluino masses at low energies, when a minimal Supergravity scenario is considered to be present at the GUT scale~\cite{Martin:1997ns}. A generalization of this procedure to underlying models different from the RPV-SUSY example is straightforward.

In Sec.~\ref{sec:Rates}, after having introduced the standard mechanism, we will discuss the actual formula for the corresponding decay rate.

\section{\label{sec:chi2}The standard analysis of a positive signal}

The most interesting case to study is the one where the GERDA experiment yields a positive signal. This corresponds to a measured decay rate $\Gamma_{\rm obs}$ that is different from zero and has a standard deviation $\sigma_{\rm obs}$:\footnote{Note that, depending on the size of $\sigma_{\rm obs}$, it is possible that this rate is also consistent with a zero decay rate at $x\cdot \sigma$ if $\Gamma_{\rm obs}- x \cdot \sigma_{\rm obs}=0$. We will ignore this subtlety here and assume that a positive experimental signal means that lepton number violation is experimentally established. This is a reasonable assumption in the likely case that we trust the experimentalists conducting the experiment. Furthermore, the analysis would become much more subtle when including such pathologic cases, without any significant gain.}
\begin{equation}
 \Gamma_{\rm obs}\pm \sigma_{\rm obs}\ @~68~\%~{\rm C.L.}
 \nonumber
\end{equation}
Assuming that we interpret this measurement as experimental proof of lepton number violation, we would, as a next step, try to describe it by assuming the standard light neutrino exchange for the mediation of the decay~\cite{Bilenky:1987ty}. In this case, our expectation for the decay rate would be~\cite{Pascoli:2005zb}
\begin{equation}
 \Gamma_\nu=G |\mathcal{M}^{0\nu}|^2 |m_{ee}|^2,
 \label{eq:0nbb-rate_nu}
\end{equation}
where $G$ is a known phase space factor, $\mathcal{M}^{0\nu}$ is the nominal nuclear matrix element (NME), and $|m_{ee}|$ is the effective neutrino mass. The latter is a function of the smallest neutrino mass $m$, the mixing angles $\theta_{12}$ and $\theta_{13}$, the mass-squared differences $\dms$ and $\dma$, as well as the two Majorana phases $\alpha$ and $\beta$.\footnote{Note that $|m_{ee}|$ does not, contrary to some statements in the literature, depend on the Dirac CP-phase $\delta$, since this phase can be rotated away by redefining the third neutrino mass eigenstate $\nu_3$, see Refs.~\cite{Lindner:2005kr,Merle:2006du}.}

To determine the goodness of fit of the measured data with the standard neutrino-exchange mechanism, we will construct a $\chi^2$-function following the procedure outlined in Ref.~\cite{Pascoli:2005zb}, extended by information on $m_\beta^2$. In that paper, the authors have used the method of the covariance matrix given by
\begin{equation}
 S_{ab}=\delta_{ab} \sigma_{a,\rm exp}^2 + \sum_i \frac{\partial T_a}{\partial x_i} \frac{\partial T_b}{\partial x_i} \sigma_i^2,
 \label{eq:covariance}
\end{equation}
where $a,b\in \{ 1,2,3\}$, $\sigma_{a,\rm exp}$ are the experimental errors of the measured observables, $x_i\in \{ s_{12}^2, s_{13}^2, \dms, \dma \}$ are the relevant neutrino oscillation parameters, and $\sigma_i\in \{ \sigma(s_{12}^2), \sigma(s_{13}^2), \sigma(\dms), \sigma(\dma) \}$ are the corresponding standard deviations. Explicitly, the $T_a$'s are given by
\begin{equation}
 T_1=\xi |m_{ee}|, \quad T_2=m_\beta^2, \quad T_3=\Sigma,
 \label{eq:covariance_quantities}
\end{equation}
where
\begin{eqnarray}
 |m_{ee}|&=&|m_1 c_{12}^2 c_{13}^2 + m_2 s_{12}^2 c_{13}^2 e^{2i\alpha} + m_3 s_{13}^2 e^{2i\beta}|,\nonumber\\
 m_\beta^2&=&m_1^2 c_{12}^2 c_{13}^2 + m_2^2 s_{12}^2 c_{13}^2 + m_3^2 s_{13}^2,\label{eq:covariance_observables}\\
 \Sigma&=&m_1+m_2+m_3.
 \nonumber
\end{eqnarray}
The above observables are the main sources of information about the neutrino mass that we currently have:
\begin{itemize}
 \item The effective neutrino mass $|m_{ee}|$ is, as just explained, measured in experiments that search for $0\nu\beta\beta$. We will take the GERDA experiment~\cite{Abt:2004yk} as an example.
 \item The effective mass square of the electron neutrino, $m_\beta^2$, is measured in kinematical experiments on single $\beta$-decay, where KATRIN~\cite{Osipowicz:2001sq} is currently the most prominent upcoming example.
 \item The effective sum $\Sigma$ of neutrino masses as measured in cosmological observations. This can be done through measurements of the cosmic microwave background (CMB) radiation, where we take the Planck satellite~\cite{Planck:2011ah} as example. However, the CMB measurements need to be combined with results from other observations such as high redshift galaxy surveys, baryon acoustic oscillations, or of Type Ia supernovae~\cite{Hannestad:2006zg,Hannestad:2007cp} to yield robust results.
\end{itemize}
Using these observables, the full $\chi^2$-function is given by
\begin{equation}
 \chi^2(m,F, \alpha, \beta)=\min\limits_{\xi \in [1/\sqrt{F},\sqrt{F}]} v^T S^{-1} v,
 \label{eq:chi2_total}
\end{equation}
where $v^T=(T_1-T_1^{\rm obs},T_2-T_2^{\rm obs},T_3-T_3^{\rm obs})$, and $T_a^{\rm obs}$ are the experimental values of the measured quantities. In particular, in the case of \onbb, we have
\begin{equation}
 T_1^{\rm obs} = |m_{ee}|_{\rm obs}= \sqrt{\frac{\Gamma_{\rm obs}}{G}}\frac{1}{|\mathcal{M}^{0\nu}|}, \quad \sigma_{1, {\rm exp}} = \sigma (|m_{ee}|) = \frac12\frac{\sigma_{\rm obs}}{\Gamma_{\rm obs} }|m_{ee}|_{\rm obs}, \quad \xi=\frac{|\mathcal{M}^{0\nu}_{\rm true}|}{|\mathcal{M}^{0\nu}|}.
 \label{eq:chi2_onbb_add}
\end{equation}
Here, $|\mathcal{M}^{0\nu}|$ is the nominal value of the NME used in the analysis to extract the experimental value of the effective mass, whereas $|\mathcal{M}^{0\nu}_{\rm true}|$ is its (unknown) true value. Furthermore, $F$ parametrizes the uncertainty in the NME, since it forces the $\chi^2$-minimization to treat all values of the NME between $|\mathcal{M}^{0\nu}|/\sqrt{F}$ and $|\mathcal{M}^{0\nu}|\sqrt{F}$ on the same footing, corresponding to a flat prior.  As representative values we will choose $F=1$, $2$, and $3$, where $F=1$ corresponds to a perfect knowledge of the NME (cf.\ Ref.~\cite{Pascoli:2005zb}).
Note that, depending on which information is available, it might also be required to reduce Eq.~\eqref{eq:chi2_total} to incorporate, e.g., $|m_{ee}|$ and $\Sigma$ only, but not $m_\beta^2$. From Eq.~\eqref{eq:chi2_total}, it is possible to derive 1$\sigma$-ranges for the smallest neutrino mass $m$.

\section{\label{sec:scenarios}The benchmark scenarios}

\begin{figure}[t!]
\centering
\begin{tabular}[ht]{lr}
\includegraphics[width=0.47\textwidth]{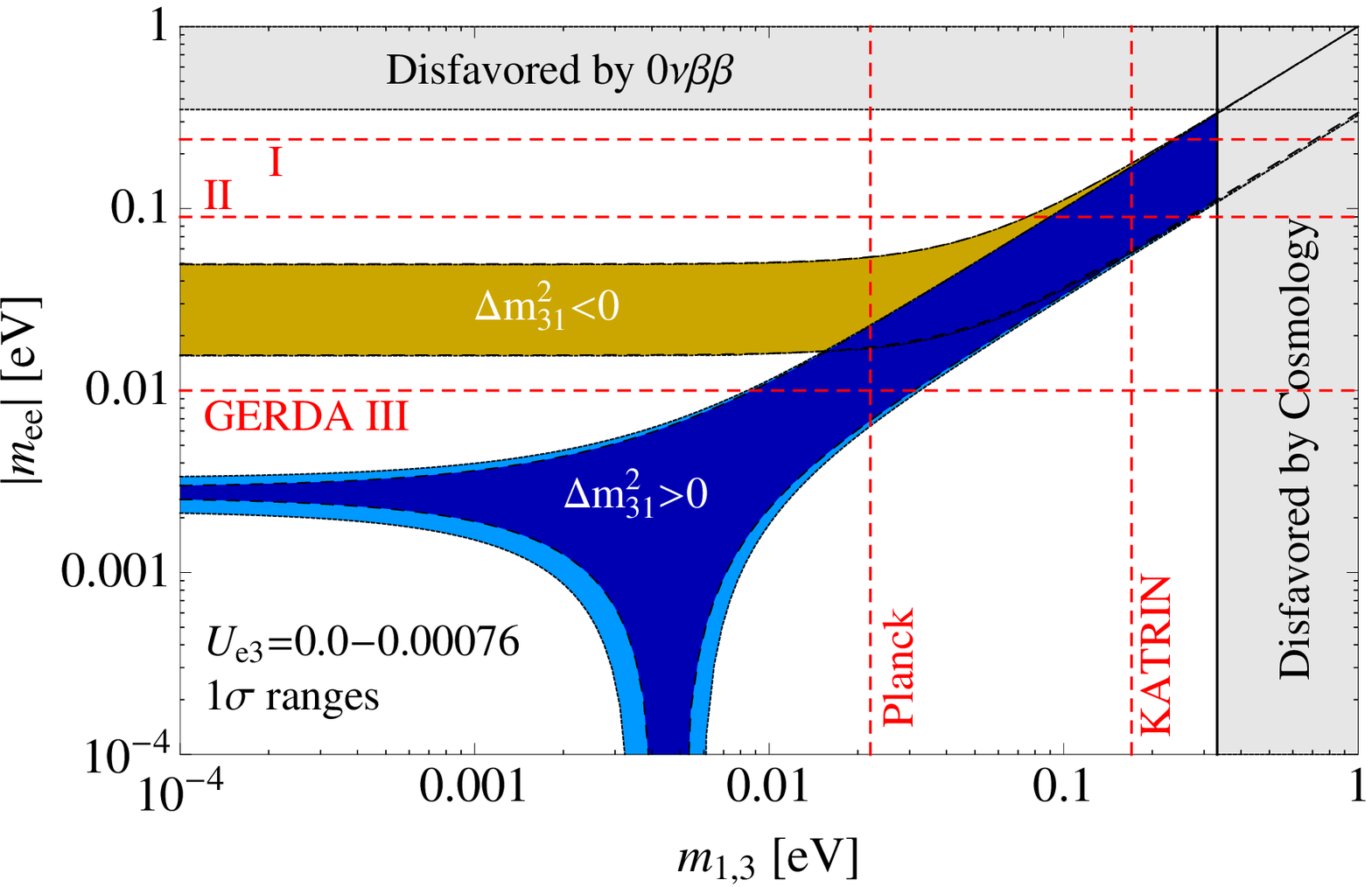}  &
\includegraphics[width=0.47\textwidth]{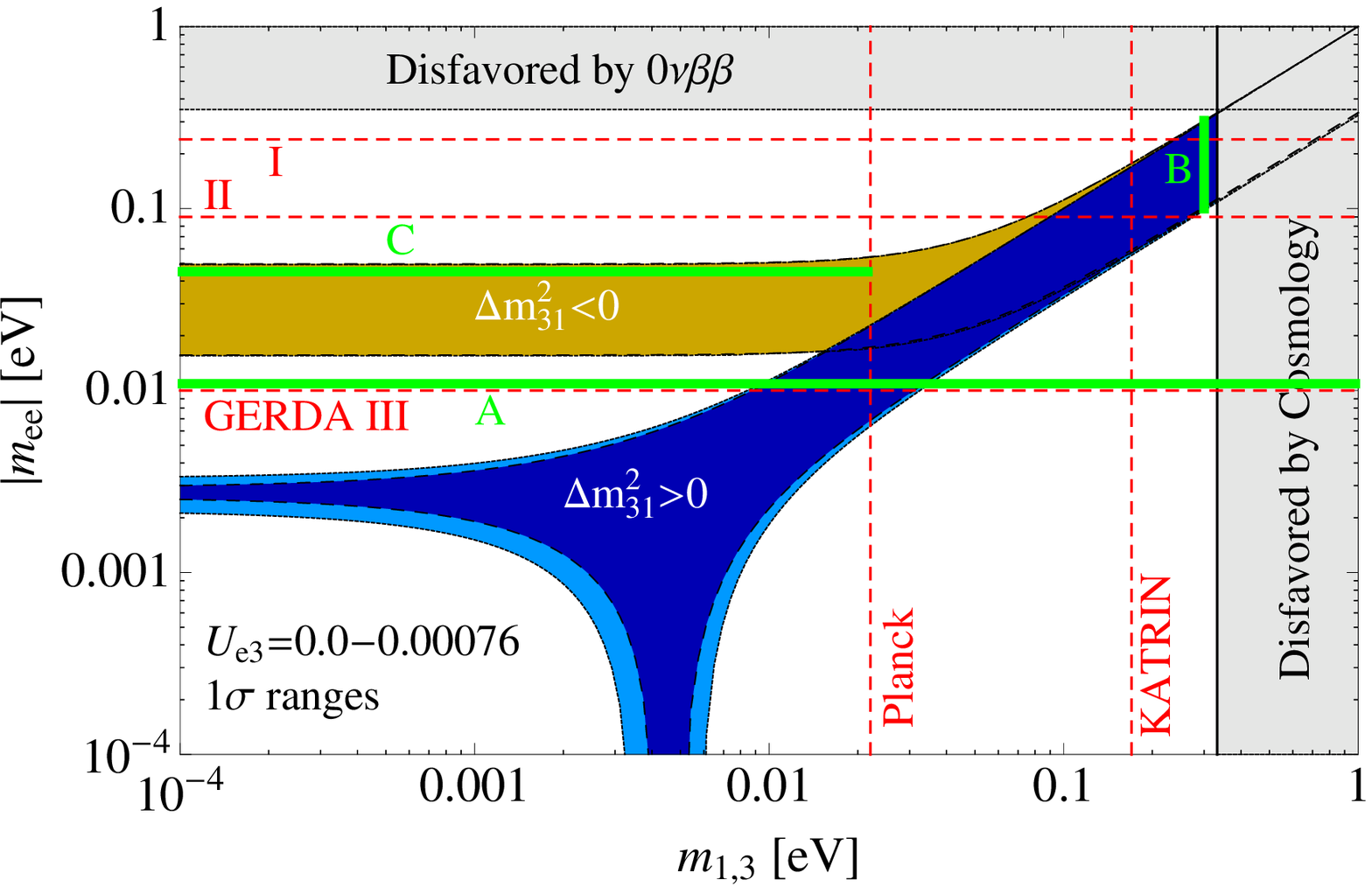}
\end{tabular}
\caption{\label{fig:scenarios} The experimental situation in a few years from now and the benchmark scenarios (A,B,C) to be considered.}
\end{figure}

The experimental situation in a few years from now is depicted in the left panel of Fig.~\ref{fig:scenarios}, which shows the range of the effective mass $|m_{ee}|$ in $0\nu\beta\beta$ as a function of the smallest neutrino mass $m$ for both orderings, normal ($m_1 < m_2 < m_3$) and inverted ($m_3 < m_1 < m_2$). The inner (outer) bands show the respective $1\sigma$ region for a value of $\sin^2 \theta_{13}=0$ (for the Double Chooz limit in case of a negative signal, $\sin^2 \theta_{13}<0.0076$~\cite{Palomares:2009wz}), while for the other oscillation parameters we have used the best-fit values and (symmetrized) standard deviations from Ref.~\cite{Schwetz:2011qt}. Obviously, the effect of a tiny but non-zero mixing angle $\theta_{13}$, is very small, in agreement with Refs.~\cite{Minakata:2001kj,Merle:2007hv}. In fact, for inverted ordering, the outer band (lighter yellow) is barely visible. The only effect of a non-zero $\theta_{13}$ shows up for the region of normal hierarchy (normal ordering with very small $m_1$), which cannot be probed in the near future. Accordingly, the dependence of the effective mass on the Majorana phase $\beta$ is very weak~\cite{Lindner:2005kr}.

We have also depicted the expected sensitivities of future experiments. First of all, the GERDA experiment will set different limits on the half-life of $0\nu\beta\beta$ in three different phases I, II, and III, which are given by $3\cdot 10^{25}$~years~\cite{Jochum:2010zz}, $1.5\cdot 10^{26}$~years~\cite{Jochum:2010zz}, and $2.8\cdot 10^{27}$~years~\cite{GERDA-Talk}. To translate these values into sensitivities on $|m_{ee}|$, one has, in principle, to know the correct NME. Since we do not want to enter a discussion about better and worse values of this quantity, we have taken, for simplicity, the most optimistic values from Ref.~\cite{Abt:2004yk} for phases I and II. As the reach of phase III is not completely clear yet (and hence, there is, to our knowledge, no official number from the GERDA collaboration), we have taken the benchmark value $|m_{ee}|=0.01$~eV, which is in agreement with recent estimates~\cite{GERDA-Talk}.

We have also depicted limits from complementary experiments that can yield information on the absolute neutrino mass scale. The only model-independent way to measure the neutrino mass is from simple kinematics, which will be performed by the KATRIN experiment~\cite{Osipowicz:2001sq}. Depending on the statistics used, one obtains different predictions for the sensitivity of the KATRIN experiment~\cite{Host:2007wh}, and we have used the slightly more optimistic value obtained by using Bayesian statistics, $m_\beta=0.17~{\rm eV}$ (which translates into just the same number for the smallest neutrino mass). The Planck experiment is supposed to probe the effective sum $\Sigma$ of light neutrino masses down to about $0.1$~eV~\cite{Hannestad:2006zg}, which we have taken as a benchmark value (and which translates into a limit on the smallest neutrino mass of roughly 0.02~eV).\footnote{Note that one might have to be careful when taking the cosmological observations at face value~\cite{Maneschg:2008sf}, since unknown systematic errors might be involved.}

Based on these numbers, we have chosen our three benchmark scenarios indicated by the solid green lines in the right panel of Fig.~\ref{fig:scenarios}, which in principle correspond to the three basic situations that are possible after the next generation of experiments:\footnote{However, note that there might of course be some subtleties involved, in particular in cases where the experimental results point to regions in parameter space which may correspond to more than one of our scenarios. We do not take into account such peculiarities, as long as there is no real measured signal which indeed falls in exactly such a region.}

\begin{enumerate}

\item[A:] {\it negative signal}\\
In this situation, no signal of \onbb\ has been measured by GERDA. No matter if we have additional information on the neutrino mass by kinematic measurements or by cosmology, we do not know if neutrinos are Dirac or Majorana particles. Correspondingly, we cannot use $|m_{ee}|$ to constrain $m_\beta$ and/or $\Sigma$, or vice versa. Concerning new physics, we can only give an upper bound on the size of the corresponding amplitude.

\item[B:] {\it consistent positive signal}\\
Here, GERDA has measured a certain rate of \onbb\, and the corresponding range obtained for $|m_{ee}|$ is consistent with complementary positive signals from KATRIN and Planck, as in the ``$\mathcal{QD}$ scenario'' of Ref.~\cite{Maneschg:2008sf}. In this case, one can fit all signals with the minimal scenario and we expect strong limits on contributions from further new physics. Depending on which term dominates, this situation will lead to a strong upper bound on $|\eps_i|$ or on the interference term $|\eps_i \cos \phi_i|$, cf.\ discussion in Sec.~\ref{sec:Analysis_B}.

\item[C:] {\it inconsistent positive signal}\\
Finally, we consider the case where GERDA has measured a non-zero rate of \onbb , which does, however, disagree with the values of $|m_{ee}|$ consistent with the measurements by KATRIN and Planck. In particular, these measurements are not consistent with a \onbb\ rate as high as measured. Then, we are forced to conclude that $0\nu\beta\beta$ is actually dominated by the contributions from new physics. In such a situation, one can estimate or even measure the magnitude of $|\epsilon_i|$.

\end{enumerate}
These three benchmark scenarios will be used in the following to derive the respective information on the contributions of point-like new physics to $0\nu\beta\beta$.

Finally, we want to remark that we could, of course, have chosen a different selection of experiments, in particular of future and upcoming searches for $0\nu\beta\beta$ (e.g.\ EXO~\cite{Danilov:2000pp,Conti:2003av}, Majorana~\cite{Gaitskell:2003zr}, or MOON~\cite{Ejiri:1999rk,Shima:2008zz}) or for the kinematical determination of the neutrino mass (e.g.\ MARE~\cite{Nucciotti:2010tx}). When having a look at the next five years, however, we consider our selection of experiments and scenarios to be both, representative and realistic.

\section{\label{sec:Rates}Half-lives and numerical analysis}

The full half-life obtained from the Lagrangian in Eq.~\eqref{eq:L_0nbb} in combination with the standard light neutrino exchange can be derived from Refs.~\cite{Pas:2000vn} and~\cite{Doi:1985dx}, and is given by
\begin{equation}
 [T_{1/2}^{0\nu\beta\beta}]^{-1}=G_1 \left| \sum_{i=0}^3 \mathcal{\tilde M}_i \right|^2 + G_2 \left| \sum_{i=4}^5 \mathcal{\tilde M}_i \right|^2+ G_3\ {\rm Re}\left[ \left( \sum_{i=0}^3 \mathcal{\tilde M}_i \right) \left( \sum_{i=4}^5 \mathcal{\tilde M}_i \right)^* \right].
 \label{eq:total_rate}
\end{equation}
The sum consists of three parts, namely the squared terms originating from the amplitudes labeled $(0,1,2,3)$, the squared terms originating from the amplitudes labeled $(4,5)$, and the interference terms.\footnote{Note that, depending on the helicities of the final state electrons, several of the interference terms in Eq.~\eqref{eq:total_rate} can be zero. This point will be discussed in a moment.} Although all operators have two electrons in the final state, the dependence of the nuclear/atomic physics parts on the final state energies are different~\cite{Doi:1985dx}, which is the reason for the appearance of the three different phase space factors $G_{1,2,3}$~\cite{Pas:2000vn}. The subscript ``0'' denotes the part coming from light neutrino exchange, while the other subscripts label the corresponding operators in Eq.~\eqref{eq:L_0nbb}. Accordingly, one has
\begin{equation}
 \mathcal{\tilde M}_i= \left\{
 \begin{matrix}
 \mathcal{M}_0 & {\rm for}\ i=0,\\
 \eps_i \mathcal{M}_i & {\rm for}\ i>0.
 \end{matrix}
 \right.
 \label{eq:mod_amps}
\end{equation}
Here, the amplitude for the light neutrino exchange is given by the standard expression~\cite{Doi:1985dx,Bilenky:1987ty,Schmitz:1997ag,Simkovic:2010ka},
\begin{equation}
 \mathcal{M}_0=\mathcal{M}^{0\nu} \frac{m_{ee}}{m_e}, \quad {\rm with}~~ \mathcal{M}^{0\nu}=-\frac{g_V^2}{g_A^2} \mathcal{M}_{F,\nu}+\mathcal{M}_{GT,\nu},
 \label{eq:light_nu_amp}
\end{equation}
and all the other NMEs $\mathcal{M}_i$ as well as the phase space factors are specified in Ref.~\cite{Pas:2000vn}.

Note that there might be CP- or other phases involved, i.e., the coefficients $\eps_i$ might effectively be complex numbers (where any overall phase will cancel out, but a relative phase can remain). Hence, it is necessary to think about how to deal with possible interference terms. Sometimes in the literature, such phases are ignored by assuming CP-invariance~\cite{Ribeiro:2007ud}, which then results in relative signs only, which could be plus or minus~\cite{Schmitz:1997ag}, leading to minimal or maximal cancellation of the partial amplitudes. We will discuss this point in the following.

First of all, one should ask the question when interference terms can appear at all. The key point is that states of a different helicity will not interfere, as they have a quantum number that distinguishes them, a fact that is well known for, e.g., $\mu^- \mu^+ \to e^- e^+$ scattering~\cite{Peskin:1995ev}. But how is the situation in a \onbb -transition? As is usual, we consider only a $0^+ \to 0^+$ nuclear transition, which is by far the dominant contribution~\cite{Doi:1985dx}. Hence, no matter which chiralities were chosen in Eq.~\eqref{eq:hadron_operators}, the decay will always have {\it the same} initial state, namely a nucleus with zero spin and positive parity. Also, the nuclear final state will have zero spin. Furthermore, since the $s$-wave approximation is taken for the final state electrons~\cite{Pas:2000vn}, they will have a zero orbital angular momentum. In such a situation, the conservation of angular momentum forces the final electron spins to point in opposite directions. However, this does not constrain the helicities of the final state electrons, since the nucleus itself can obtain a non-zero 3-momentum, which allows all electron helicities to be produced.\footnote{In other words: The correct helicities are enforced by the electrons being emitted in certain directions. However, these directions cannot be detected in an experiment like GERDA.} Accordingly, even though we start with the same initial state for all transitions, we can have {\it different} final states, which will {\it not} interfere.

What we can learn from this discussion is that the decisive point is {\it only} the helicities of the final state {\it electrons}, cf.\ Eq.~\eqref{eq:electron_operators}. However, the final state quarks do not matter (exactly as the initial state quarks do not matter), since they are ``hidden'' inside the nucleus, and their helicities are never measured. Hence, even though the initial or final state helicities in Eq.~\eqref{eq:hadron_operators} might differ on the quark level, it does not matter for the actual decay, since the quarks are only intermediate states.

This leads us to a question: Which of the operators in Eq.~\eqref{eq:L_0nbb} can actually have the same final state electron helicities as the standard light neutrino exchange process? This is easy to answer: Using the symmetries involved, one can see that, in the standard case, the electron current is given by $j_R$ from Eq.~\eqref{eq:electron_operators}, due to the double weak vertex on the Majorana neutrino line~\cite{Bilenky:1987ty}. This leads to two left-handed electrons in the final state, as to be expected from the left-handedness of the weak interactions.\footnote{Actually, we are sloppily setting helicity and chirality equal, which is strictly speaking only justified in the highly relativistic limit. Nevertheless, the kinetic energy of both electrons for $^{76}$Ge will be $2.04$~MeV~\cite{Schmitz:1997ag}, making them relativistic, but not highly relativistic. Although this might lead to some error, we have to take into account that the largest error will actually come from the problematic determination of the NMEs involved.}

We can conclude: The only operators that can interfere with the standard process are the ones with the coefficients $\eps_1$, $\eps_2$, and $\eps_3$, and even they can only interfere if the current $j_R$ is involved. Accordingly, one can constrain the following operator coefficients without having to care about interferences: $\eps_{1}^{xyL}$, $\eps_{2}^{xyL}$, $\eps_{3}^{xyL}$, $\eps_{4}^{xyz}$, and $\eps_{5}^{xyz}$, where $x,y,z \in \{ L,R \}$.

Note that interference effects between different modes of $0\nu\beta\beta$ have also been discussed in Refs.~\cite{Faessler:2011qw,Faessler:2011rv}. They both agree with our findings in that the important condition for the appearance of interference effects is the equality in the chiral structures of the new physics contribution and the light neutrino exchange diagram, and hence the equality of the corresponding phase space factors. Furthermore, Ref.~\cite{Faessler:2011rv} points out that the introduction of new physics contributions could easily lead to relative CP-phases, as we have also mentioned.

Finally, we want to note that interferences will, if they appear at all, only play a role in scenario~B. The reason is simply that for scenarios~A and~C, it is perfectly enough to consider the contribution from the respective effective operator only.

These considerations allow us to write down expressions for the half-lives for all three benchmark scenarios:
\begin{enumerate}

\item[A:] {\it negative signal}\\
Here, only the contribution from the effective operators can be constrained. The expression we need is
\begin{equation}
 [T_{1/2}^{0\nu\beta\beta}]^{-1}_{{\rm A},i}=\tilde G_i |\eps_i|^2 \left| \mathcal{M}_i \right|^2,
 \label{eq:T12_A}
\end{equation}
where $i=1,2,3,4,5$, $\tilde G_{1,2,3}=G_1$, and $\tilde G_{4,5}=G_2$.

\item[B:] {\it consistent positive signal}\\
In this case, we will also need to consider the contribution coming from the standard light neutrino exchange and obtain
\begin{eqnarray} \label{eq:T12_B}
[T_{1/2}^{0\nu\beta\beta}]^{-1}_{{\rm B},i} &=& G_1 \left| \mathcal{M}_0 \right|^2\ +\ \tilde G_i |\eps_i|^2 \left| \mathcal{M}_i \right|^2   \\
& & \left( +\ 2 G_1 \cos \phi_i \cdot |\eps^{xyR}_i| \left| \mathcal{M}_0 \right| \left| \mathcal{M}_i \right|\ {\rm for}\ i=1,2,3\right), \notag
\end{eqnarray}
where $i=1,2,3,4,5$, and $\phi_i\equiv \phi^{xyR}_i$ is the relative phase between the two different contributions in the relevant cases. Of course, we cannot give an explicit expression for $\phi_i$ as long as the underlying theory is unknown, but e.g.\ in the case of the exchange of new heavy active neutrinos, this would correspond to a new Majorana phase $\gamma$ in addition to the two phases $\alpha$ and $\beta$ for light neutrinos.

\item[C:] {\it inconsistent positive signal}\\
Also here, we only need the contribution from the effective operators, which yields the same formula as for scenario A,
\begin{equation}
 [T_{1/2}^{0\nu\beta\beta}]^{-1}_{{\rm C},i}=\tilde G_i |\eps_i|^2 \left| \mathcal{M}_i \right|^2.
 \label{eq:T12_C}
\end{equation}

\end{enumerate}
These scenarios will now be analyzed numerically.

\subsection{\label{sec:Analysis_A}Scenario A: Negative signal}

This scenario is the easiest one to analyze. The key point is that, in the case of a negative signal, we cannot use any external information on the neutrino parameters. The reason is that, even though we could in principle calculate a range for the effective mass $|m_{ee}|$ from data obtained by neutrino oscillation experiments in combination with kinetic and/or cosmological neutrino mass measurements, lepton number violation would not have been experimentally established, and hence, we would not even know if $|m_{ee}|$ has any physical meaning. Neutrinos could be Dirac particles, still in agreement with all experimental results to date.\footnote{Of course, this will change if we know from any other type of experiment that lepton number violation exists and that neutrinos are Majorana fermions. However, such a situation is not very likely to be present before data from the GERDA experiment will be available~\cite{Merle:2006du}, so we do not consider this case here.}

The only decent statement we can make in such a situation is that the contribution $\Gamma_\epsilon$ to \onbb\ from effective point-like operators can be at most as large as the upper bound on the decay rate measured in the experiment, i.e.,
\begin{equation}
 \Gamma_\epsilon < \Gamma_{\rm bound}.
 \label{eq:eps-range_negative}
\end{equation}
Note that this limit is a conservative one in the sense that we allow the contribution from the effective operators to be the dominant one, while any possible interference terms or further contributions are neglected. One could, of course, improve the bounds by assuming additional contributions rendering the point-like interactions much less dominant. However, the corresponding limit would only hold under this assumption and would hence be much more model-dependent.

Furthermore, Eq.~\eqref{eq:eps-range_negative} should be viewed as an estimate rather than as a strict bound at a certain confidence level, due to the intrinsic uncertainties associated with the computation of the NMEs. We will comment on this point in Sec.~\ref{sec:Caveat}.

In our analysis, we will always assume a single term from Eq.~\eqref{eq:L_0nbb} with a certain chirality structure to be dominant\footnote{This is what is called ``on-axis evaluation'' in Ref.~\cite{Pas:2000vn}.}, e.g.\ $\epsilon_5^{LRL} J_L^\mu J_R j_{L,\mu}$. This will lead to a certain decay rate $\Gamma_{\epsilon_5^{LRL}}$ that is a function of the corresponding coefficient $\epsilon_5^{LRL}$. Applying Eq.~\eqref{eq:eps-range_negative} will then yield an upper bound on $\epsilon_5^{LRL}$ .

A similar analysis has been done in Ref.~\cite{Pas:2000vn} for a 90~\%~C.L.\ limit obtained by the Heidelberg-Moscow experiment ($T_{1/2}^{0\nu\beta\beta}>1.8\cdot 10^{25}$~years, Refs.~\cite{Pas:2000vn,Baudis:1999xd}), which we have also reproduced as a cross check, using the same numbers as in the original reference.\footnote{Note that we could reproduce all numbers from Ref.~\cite{Pas:2000vn} except for the bound on $|\eps_3^{LLz,RRz}|$, where our value is smaller by a factor of two.}

To adjust the numbers to our scenario A, we have taken the 90~\%~C.L.\ limit for GERDA phase III in the case of a negative signal, $(T_{1/2}^{0\nu\beta\beta})_{\rm A}>2.8\cdot 10^{27}$~years~\cite{GERDA-Talk}. The nuclear radius for $^{76}$Ge has been estimated as $R \simeq 1.2 \sqrt[3]{76}$~fm $\simeq 5.1$~fm, and the numerical values for the phase space factors have been taken from Ref.~\cite{Pas:2000vn}. However, it was necessary to update the values of the NMEs, since there have been many developments within the last years, in particular in what regards the inclusions of short-range correlations and higher order nuclear currents~\cite{Fedor}. Taking the calculations from Refs.~\cite{Simkovic:2010ka,Fedor}, we can update the values from Ref.~\cite{Pas:2000vn} to be\footnote{Note that the matrix elements used in Ref.~\cite{Pas:2000vn} were based on the calculations from Ref.~\cite{Hirsch:1995ek}, up to a different sign convention in the definition of the Fermi part. The values from Refs.~\cite{Simkovic:2010ka,Fedor}, however, were based on the calculations from Ref.~\cite{Simkovic:1999re}, which uses a normalization for the individual Fermi and the Gamow--Teller parts that is smaller by a factor of $m_e/m_p$, but the full NME is of course of the same order of magnitude.}
\begin{equation}
 \mathcal{M}_{F,N}=6.8\cdot 10^{-2} \quad {\rm and} \quad \mathcal{M}_{GT,N}=1.7\cdot 10^{-1}.
 \label{eq:NME_updates}
\end{equation}
Note that we also use $g_A=1.25$, as suggested in Ref.~\cite{Simkovic:2010ka}. In addition, all other numerical values are taken from Ref.~\cite{Pas:2000vn}.
\begin{table}[t]
\centering
\begin{tabular}{|c||c|c|c|c|c|c|}\hline
A & $|\eps_1|$ & $|\eps_2|$ & $|\eps_3^{LLz,RRz}|$ & $|\eps_3^{LRz,RLz}|$ & $|\eps_4|$ & $|\eps_5|$\\ \hline \hline
Bound & $1.6\cdot 10^{-8}$ & $9.5\cdot 10^{-11}$ & $1.2\cdot 10^{-9}$ & $7.3\cdot 10^{-10}$ & $7.9\cdot 10^{-10}$ & $7.1\cdot 10^{-9}$\\ \hline
\end{tabular}
\caption{\label{tab:results_A} The resulting upper bounds on the magnitudes of the coefficients of the effective operators in scenario A, where $z=L$ or $R$.}
\end{table}

The resulting upper bounds on the magnitudes of the coefficients of the effective operators are given in Tab.~\ref{tab:results_A}, where we have indicated the chirality structures only for the cases where the bounds depend on them. As to be expected for scenario A, the bounds on the coefficients will be stronger than the bounds based on the Heidelberg-Moscow limit by about one order of magnitude, since phase III of GERDA would yield a limit on the decay rate that is about two orders of magnitude better, and since the squared absolute values of the coefficients enter the decay rate.

Finally, we would like to also derive a bound on the paremeter $\lambda'_{111}$ in our RPV-SUSY example. Using Eq.~\eqref{eq:eps_RPV}, as well as the bound on $\epsilon_1^{RRR}$ from Tab.~\ref{tab:results_A}, it is easy to derive the value $\lambda'_{111}<0.06$ for the sparticle masses assumed, which is a reasonable value as compared to other bounds on this parameter~\cite{Chemtob:2004xr}.

\subsection{\label{sec:Analysis_B}Scenario B: Consistent positive signal}

In case of a positive signal from GERDA, one first needs to check if all the available data from all experiments is consistent with the assumption that light neutrino exchange is the only mechanism behind $0\nu\beta\beta$. We will also investigate situations in which only {\it one} experiment (KATRIN or Planck) can yield complementary information on the neutrino mass, which would be important if one of the two ceases to produce reliable results.

The main technical problem is how to separate the point-like contributions resulting from the interactions in Eq.~\eqref{eq:L_0nbb} from the dominant contribution of light neutrino exchange, which is a long-range force~\cite{Pas:1997fx}. Note that, in principle, there might also be further long-range contributions, such as new heavy $W$-bosons in addition to the ones in the SM, in which case the neutrino line would still be of long range. However, we do not consider such contributions here, but instead focus on the point-like ones that can be fully expressed by effective field theory, in order to be as model-independent as possible.

To perform the analysis, we first need to assume a certain measured rate for scenario~B. Taking the smallest neutrino mass to be $0.3$~eV (cf.\ Sec.~\ref{sec:scenarios}), we roughly obtain (by varying the neutrino oscillation parameters and the Majorana phases) an effective mass $|m_{ee}|$ between 0.1~eV and 0.3~eV, which translates into a half-life between $2.8\cdot 10^{25}$~years and $2.6\cdot 10^{26}$~years for an NME of $4.0$. For definiteness, we assume a certain value within this range to be observed, let us say
\begin{equation}
 \left(T_{1/2}^{0\nu\beta\beta}\right)_{\rm B}\simeq 5.0\cdot 10^{25}~{\rm years}.
 \label{eq:TBas}
\end{equation}
This is equivalent to a decay rate of
\begin{equation}
 \Gamma_{\rm obs}= \frac{\ln 2}{\left(T_{1/2}^{0\nu\beta\beta}\right)_{\rm B}}\simeq 1.4\cdot 10^{-26}/{\rm year}.
 \label{eq:GammaBas}
\end{equation}
This measurement will, of course, have a certain error, the estimate of which can be found in Ref.~\cite{Abt:2004yk} to be
\begin{equation}
 \frac{\sigma_{\rm obs}}{\Gamma_{\rm obs}}\simeq 23~\%,
 \label{eq:GERDA_acc}
\end{equation}
which leads to an absolute error on the decay rate of $\sigma_{\rm obs}=0.32\cdot 10^{-26}/{\rm year}$. Finally, comparing Eqs.~\eqref{eq:0nbb-rate_nu}, \eqref{eq:total_rate}, and~\eqref{eq:light_nu_amp}, one observes that $G=G_1 \ln 2 / m_e^2$.
This enables the calculation of the parameters to be used in the $\chi^2$-function in Eq.~\eqref{eq:chi2_total}, namely, from Eq.~(\ref{eq:chi2_onbb_add}),
\begin{equation}
|m_{ee}|_{\rm obs} \simeq 0.23 ~{\rm eV}, \quad \sigma (|m_{ee}|)\simeq 0.026~{\rm eV},
 \label{eq:sig0nbb}
\end{equation}
where a representative nominal value of $|\mathcal{M}^{0\nu}|=4.0$ has been used for the NME~\cite{Abt:2004yk,Simkovic:2010ka}.
Consistent values for the observables measured by KATRIN and Planck can be found in Ref.~\cite{Maneschg:2008sf} for a smallest neutrino mass of $0.3$~eV: $(m_\beta^2)_{\rm obs}=(0.30~{\rm eV})^2$ and $\Sigma_{\rm obs}=0.91$~eV, respectively. Realistic values for the corresponding errors are given by $\sigma(m_\beta^2)=0.025~{\rm eV}^2$~\cite{Osipowicz:2001sq,Host:2007wh} and $\sigma(\Sigma)=0.05$~eV~\cite{Hannestad:2007cp}. For definiteness, we express all neutrino masses by the lightest neutrino mass $m$ using the formulas that hold for an {\it inverted} mass ordering. However, as we are with $m=0.3$~eV in the quasi-degenerate region of the neutrino masses, it will anyway not make much of a difference which ordering is used~\cite{Lindner:2005kr}. Finally, the values and uncertainties of the neutrino oscillation parameters we use as input are given in Tab.~\ref{tab:oscillation_params}.
\begin{table}[t]
\centering
\begin{tabular}{|c||c|c|}\hline
Parameter & Best-fit value \& 1$\sigma$ range & Reference\\ \hline \hline
$s_{12}^2$ & $0.32\pm 0.016$ & \cite{Schwetz:2011qt}\\ \hline
$s_{13}^2$ & $0.00\pm 0.0076$ & \cite{Palomares:2009wz}\\ \hline
$\Delta m_{21}^2$ & $(7.6\pm 0.19)\cdot 10^{-5}~{\rm eV}^2$ & \cite{Schwetz:2011qt}\\ \hline
$|\Delta m_{31}^2|_{\rm nor.}$ & $(2.5\pm 0.09)\cdot 10^{-3}~{\rm eV}^2$ & \cite{Schwetz:2011qt}\\ \hline
$|\Delta m_{31}^2|_{\rm inv.}$ & $(2.3\pm 0.10)\cdot 10^{-3}~{\rm eV}^2$ & \cite{Schwetz:2011qt}\\ \hline
\end{tabular}
\caption{\label{tab:oscillation_params} The best-fit values and (symmetrized) 1$\sigma$ ranges for the neutrino oscillation parameters used in the analysis. Note that, in the inverted hierarchy case, $|m_{ee}|$ is actually proportional to $\sqrt{|\Delta m_{31}^2|_{\rm inv.}}$ and depends strongly on $\theta_{12}$, whereas there is only a very weak dependence on $\theta_{13}$~\cite{Lindner:2005kr}. Accordingly, the former two observables would require a precise experimental determination in order to maximize the use of our analysis.}
\end{table}

Now we are ready to investigate the $\chi^2$-function, Eq.~\eqref{eq:chi2_total}, which will only be a function of the absolute neutrino mass scale (or, equivalently, the smallest neutrino mass $m$), the two Majorana phases $\alpha$ and $\beta$ (where the dependence on the latter phase is rather weak), and the accuracy $F$ to which we know the NME. As already mentioned, we investigate nine cases, namely the one where both KATRIN and Planck yield a result as well as the two scenarios in which only one of these experiments yields a useful measurement, and each of these three cases will be investigated for perfect knowledge of the NME ($F=1$) as well as for different degrees of imperfect knowledge ($F=2,3$).
\begin{figure}[t]
\centering
\begin{tabular}{lr}
\includegraphics[width=0.43\textwidth]{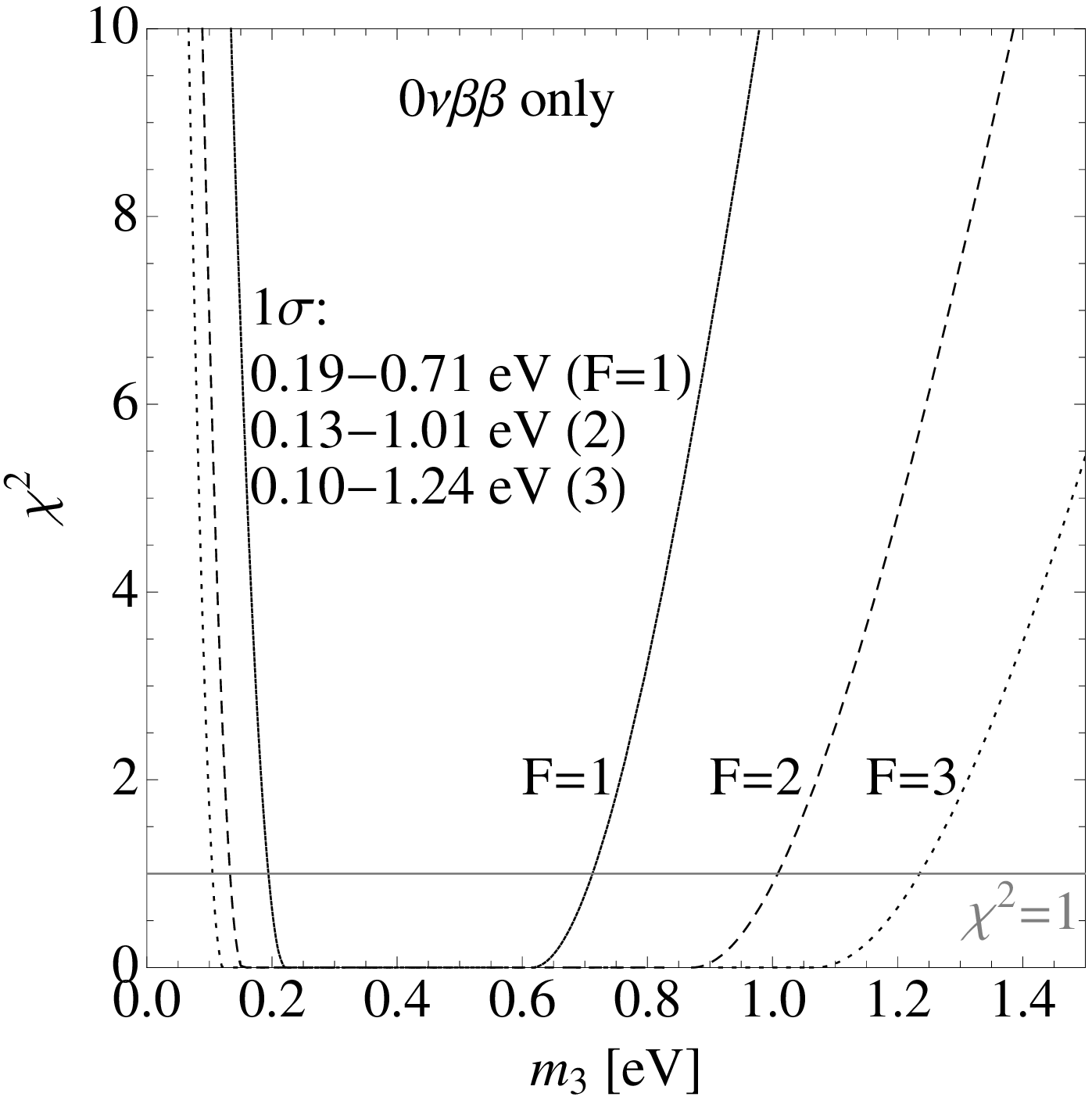}  &
\includegraphics[width=0.43\textwidth]{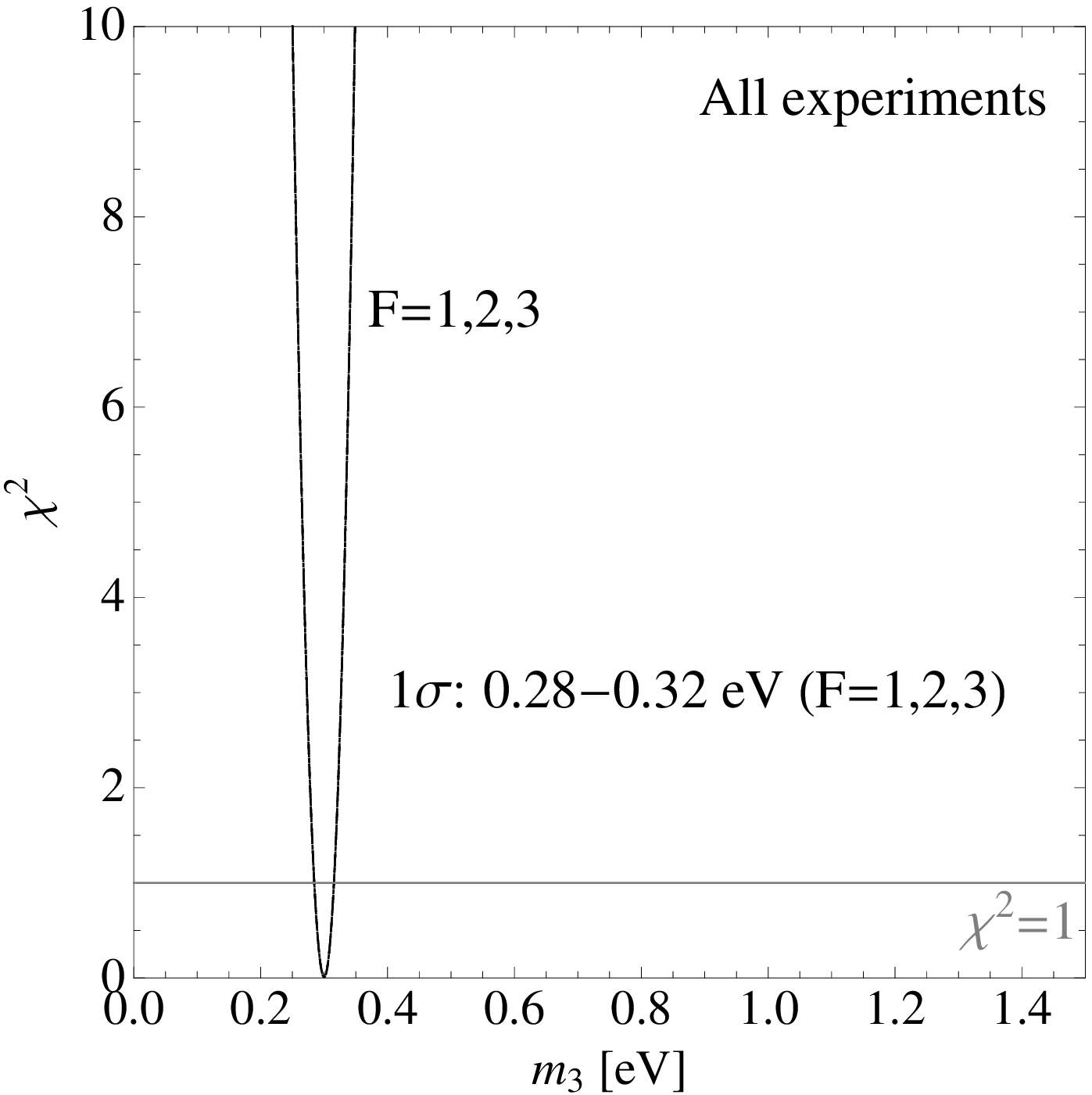}\\
\includegraphics[width=0.43\textwidth]{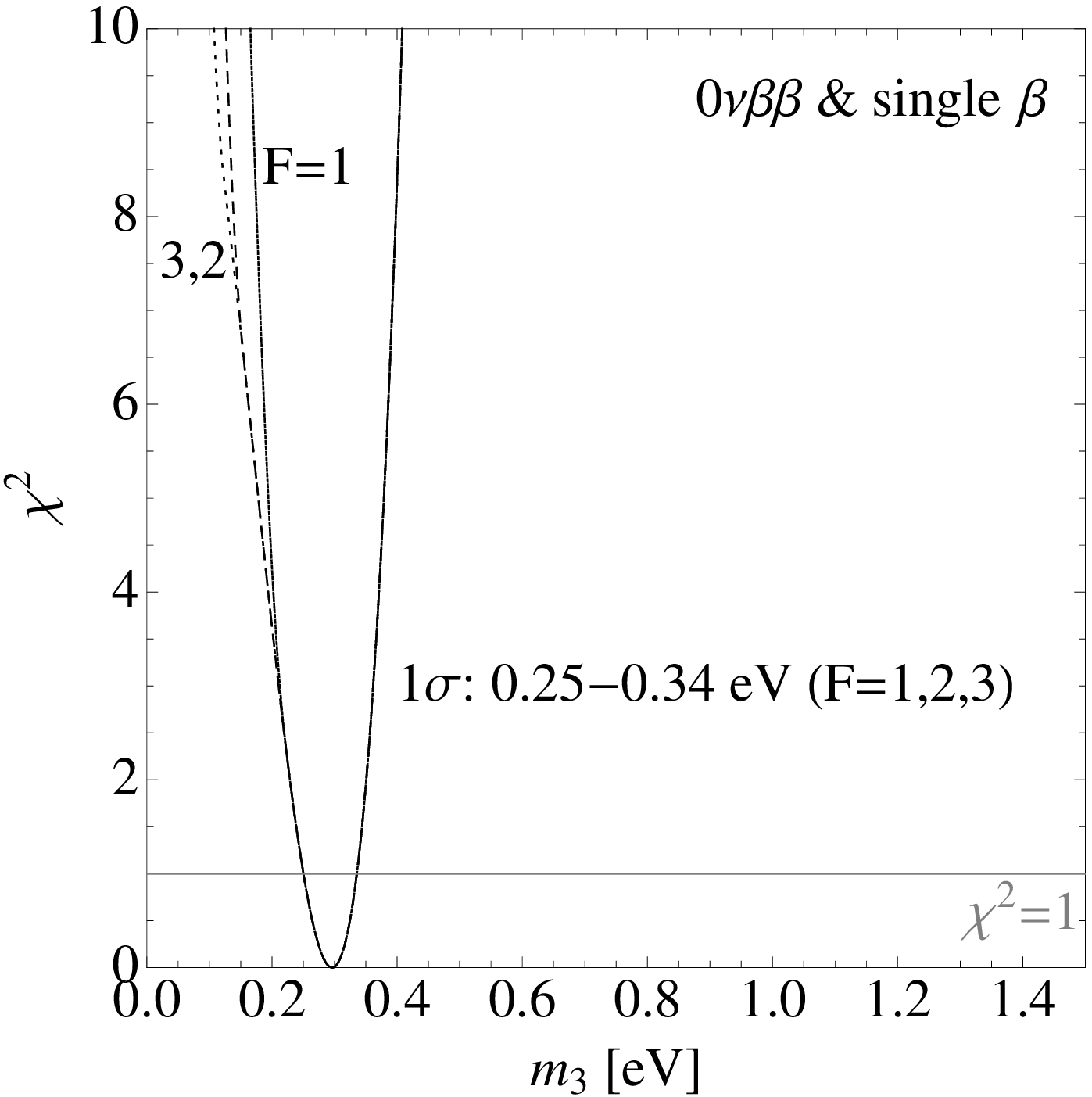}  &
\includegraphics[width=0.43\textwidth]{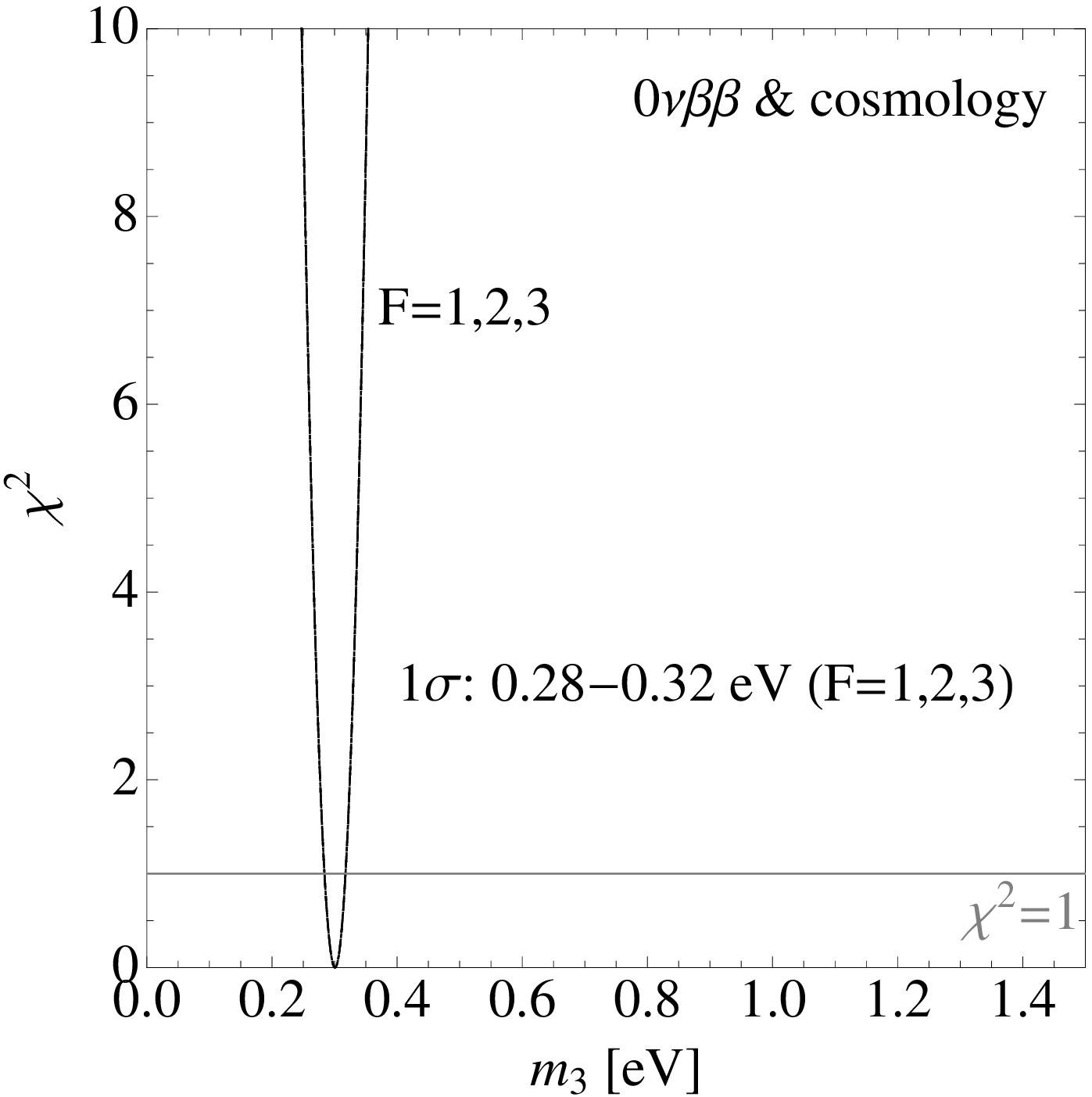}
\end{tabular}
\caption{\label{fig:chi2} The ranges for the smallest neutrino mass $m_3$ for different cases.}
\end{figure}

Following the procedure from Ref.~\cite{Pascoli:2005zb}, we minimize the $\chi^2$-function with respect to $\alpha$, $\beta$, and $\xi$. The results are the functions of $m$ shown in Fig.~\ref{fig:chi2}, where we have plotted the behavior of the $\chi^2$-function for the case where only GERDA yields a result (pessimistic), as well as for the cases where also KATRIN and/or Planck yield results. Note that a measurement of \onbb\ only (upper-left panel) does not yield an unambiguous result for the neutrino mass scale. This is natural because of the intrinsic uncertainties involved, i.e., bad knowledge of the NME involved can, in certain regions of the parameter space, be compensated by a variation of the Majorana phase $\alpha$ (and $\beta$, although this dependence is subdominant). Hence, the flat prior for the exact value of the NME corresponds to a perfectly flat region in the $\chi^2$-function.\footnote{Note that we display the full $\chi^2$-function, rather than only $\Delta \chi^2$, since our assumption of consistent results always allows for (at least) one perfect fit point.} This is different in the case when all experiments yield information (upper-right panel): Then, KATRIN as well as Planck will point to a certain neutrino mass scale, and hence, there will be no flat region left in the $\chi^2$-plot, which means that we actually have a measurement of the neutrino mass, and we can determine the corresponding best-fit point as well as the 1$\sigma$ region. The qualitative behavior does not change when we have only one experiment in addition to the \onbb-signal: If there is only the kinematical determination (lower-left panel), the $\chi^2$-function becomes broader, but it is nevertheless possible to derive a consistent range for the neutrino mass. The case of a cosmological observation being the only information in addition to \onbb\ (lower-right panel) looks practically identical to the case where all measurements fit together. This is simply a reflection of the small uncertainty assigned to the observation of the sum $\Sigma$ of neutrino masses. However, note that there could be systematic errors involved in the cosmological data that we are not aware of, which could even lead to wrong conclusions about the actual value of the neutrino mass~\cite{Maneschg:2008sf}, which is why this scenario might be a little too optimistic in case there is no additional information by KATRIN.

We have taken into account different values of the degree of knowledge of the NME ($F=1,2,3$) in all plots. However, for the 1$\sigma$ region, this only plays a role for the (too) pessimistic scenario of having a positive signal from \onbb\ only, even though the neutrino mass would be in a range where KATRIN and Planck could both determine it. We will not consider this case further.

Using the $\chi^2$-functions obtained, one can derive the best-fit values and 1$\sigma$ errors for $m$ in all cases, shown in Tab.~\ref{tab:massfit_B}. As to be expected, we always obtain our assumed value of $m=0.3$~eV as best-fit point, which is a reflection of the fact that we consider a situation in which all experiments yield consistent results. Note that, for the $1\sigma$-regions, the uncertainty in the NME does not play a major role, since we have enough complementary information from the kinematical and from the cosmological determination of the neutrino mass. Furthermore, note that, although it seems that the KATRIN experiment does not yield much more information compared to using the cosmological data alone, it nevertheless cannot be overemphasized that experiments based on kinematics are the only {\it model-independent} probe for the absolute neutrino mass scale that we have. Depending on the outcome of the cosmological observations, it might be necessary to dismiss the data on $\Sigma$ completely, in case that inconsistencies in our cosmological model would arise.
\begin{table}[t]
\centering
\begin{tabular}{|c||c|c|c|}\hline
 $F$  & KATRIN \& Planck & KATRIN only & Planck only \\ \hline \hline
 $1,2,3$  & $0.30^{+0.015}_{-0.016}$~eV & $0.30^{+0.040}_{-0.046}$~eV & $0.30^{+0.017}_{-0.017}$~eV\\ \hline
\end{tabular}
\caption{\label{tab:massfit_B} The reconstructed values and 1$\sigma$ ranges for the smallest neutrino mass.}
\end{table}

Now we are finally in the position to derive bounds on the different $|\epsilon_i|$'s. Let us shortly summarize what we have done: First of all, we started with Eq.~\eqref{eq:T12_B}, where we have neglected the interference terms, as discussed at the beginning of Sec.~\ref{sec:Rates},
\begin{equation}
 [T_{1/2}^{0\nu\beta\beta}]^{-1}_{{\rm B},i}= G_1 \left| \mathcal{M}_0 \right|^2\ +\ \tilde G_i |\eps_i|^2 \left| \mathcal{M}_i \right|^2.
 \label{eq:T12_B_later}
\end{equation}
Next, we have assumed a certain observed rate, Eq.~\eqref{eq:GammaBas}, with an error given by $\sigma_{\rm obs}$ in Eq.~\eqref{eq:GERDA_acc}.
Our goal is to derive bounds on the $|\epsilon_i|$'s, so we should re-arrange Eq.~\eqref{eq:T12_B_later} to obtain
\begin{equation}
 \tilde G_i |\eps_i|^2 \left| \mathcal{M}_i \right|^2=[T_{1/2}^{0\nu\beta\beta}]^{-1}_{{\rm B},i}- G_1 \left| \mathcal{M}_0 \right|^2=\frac{\Gamma_{\rm obs}-\Gamma_\nu}{\ln 2}.
 \label{eq:T12_B_rearranged}
\end{equation}
Since, as we have shown, the measurement of the neutrino mass scale is consistent with that of GERDA, the whole decay rate could be attributed to the standard mechanism of light neutrino exchange, causing the right-hand side of Eq.~\eqref{eq:T12_B_rearranged} to vanish.  However, the effective operators could still contribute within the error of the measurement, giving the estimate
\begin{equation}
 \tilde G_i |\eps_i|^2 \left| \mathcal{M}_i \right|^2 < \frac{\sigma_{\rm obs}}{\ln 2}.
 \label{eq:T12_B_bound}
\end{equation}
Finally, Eq.~\eqref{eq:T12_B_bound} can be rewritten to yield a bound on $|\epsilon_i|$:
\begin{equation}
 |\eps_i| < \frac{1}{\left| \mathcal{M}_i \right|} \sqrt{ \frac{\sigma_{\rm obs}}{\tilde G_i\ \ln 2} }.
 \label{eq:epsi_B_bound}
\end{equation}
\begin{table}[t]
\centering
\begin{tabular}{|c||c|c|c|c|c|c|}\hline
B & $|\eps_1| \ (!)$ & $|\eps_2| \ (!)$ & $|\eps_3^{LLz,RRz}| \ (!)$ & $|\eps_3^{LRz,RLz}| \ (!) $ & $|\eps_4|$ & $|\eps_5|$\\ \hline \hline
Bound & $5.8\cdot 10^{-8} $ & $3.5\cdot 10^{-10}$ & $4.5\cdot 10^{-9}$ & $2.7\cdot 10^{-9}$ & $2.9\cdot 10^{-9}$ & $2.6\cdot 10^{-8}$\\ \hline
\end{tabular}
\caption{\label{tab:results_B} The resulting upper bounds on the magnitudes of the coefficients of the effective operators in scenario B, where $z=L$ or $R$. Here, ``$(!)$'' denotes that there is a possibility of interference if the current $j_R$ is involved in the operator.}
\end{table}

The numerical results can be found in Tab.~\ref{tab:results_B}. One observes that the constraints for this scenario appear to be slightly worse than the ones from scenario~A (cf.\ Sec.~\ref{sec:Analysis_A}). However, we can see in Fig.~\ref{fig:scenarios} that, for scenario~A, we would actually need a successful phase~III of GERDA, while a situation like scenario~B could even be reached within phase~I, the sensitivity of which is worse by a factor of roughly 100. Since the decisive quantity for the bounds is the error rather than the decay rate itself, cf.\ Eq.~\eqref{eq:epsi_B_bound}, the bounds obtained would be strong even for phase~I. However, this is only possible if there is no inconsistency in the experimental results, cf.\ Sec.~\ref{sec:Analysis_C}.

Using the example from Eq.~\eqref{eq:eps_RPV} again, we now arrive at $\lambda'_{111}<0.12$, which is slightly worse than the value from Sec.~\ref{sec:Analysis_A}, due to the difference on the respective bounds (cf.\ Tabs.~\ref{tab:results_A} and~\ref{tab:results_B}).

One last important remark is that some of our results, i.e., those marked with ``$(!)$'' in Tab.~\ref{tab:results_B}, could be altered by the presence of an interference term. However, as already discussed, these terms will only play a role when two conditions are simultaneously fulfilled:
\begin{enumerate}

 \item the effective operator contains the electron current $j_R$ (otherwise there can be no interference in the highly relativistic limit), and

 \item the relative phase $\phi$ between the standard operator and the new physics contribution is not close to $\pi/2$.

\end{enumerate}
The first condition is fulfilled for only a handful operators [cf.~Eqs.~\eqref{eq:L_0nbb} and~\eqref{eq:electron_operators}], and even then there is still the possibility of cancellation due to the value of the relative phase (second condition). Nevertheless, we also give the bounds for such a situation. In those cases where interferences might be relevant, we expect that, unless $\cos \phi_i $ is very small in Eq.~\eqref{eq:T12_B}, the last term will dominate over the second one. Then, once again assuming that the standard mechanism of light neutrino exchange is the main contribution, one can calculate an approximate value for $|\mathcal{M}_0|$ and then derive bounds on the combination of parameters $|\eps^{xyR}_i \cos \phi_i|$ as
\begin{equation}
|\eps^{xyR}_i \cos \phi_i| < \frac12 \sqrt{\frac{\sigma_{\rm obs}}{\Gamma_{\rm obs}}}  B_i \simeq 0.24  B_i, \quad i=1,2,3,
\end{equation}
where $B_i$ is the left-hand side of Eq.~(\ref{eq:epsi_B_bound}), the values of which are given in Tab.~\ref{tab:results_B}. The resulting bounds on $|\eps^{xyR}_i \cos \phi_i|$  are shown in Tab.~\ref{tab:results_B_int}. These bounds are seemingly stronger by a factor of roughly four than the ones without interferences, but one has to keep in mind that the cosines involved may very well be of the order of 0.1, which would, in turn, weaken the bounds on the $|\eps^{xyR}_i|$'s.

Here, the RPV-SUSY parameter can only be constrained in connection with a CP-phase $\phi_\lambda$~\cite{Faessler:2011rv}. The corresponding bound from Tab.~\ref{tab:results_B_int} translates into $|\lambda'_{111} \cos \phi_\lambda|<0.028$, which could turn out to be particularly strong in case the value of the CP-phase is predicted in a certain model.

\begin{table}[t]
\centering
\begin{tabular}{|c||c|c|c|c|}\hline
B (interf.) & $|\eps^{xyR}_1 \cos \phi_1|$ & $|\eps^{xyR}_2 \cos \phi_2|$ & $|\eps^{LLR,RRR}_3 \cos \phi_3|$ & $|\eps^{LRR,RLR}_3 \cos \phi_3| $ \\ \hline \hline
Bound & $1.4\cdot 10^{-8} $ & $8.4\cdot 10^{-11}$ & $1.1\cdot 10^{-9}$ & $6.5\cdot 10^{-10}$ \\ \hline
\end{tabular}
\caption{\label{tab:results_B_int} The resulting upper bounds on the combination of parameters $|\eps^{xyR}_i \cos \phi_i|$ in scenario B, assuming that the interference terms dominate. Here, $x,y=L$ or $R$.}
\end{table}

\subsection{\label{sec:Analysis_C}Scenario C: Inconsistent positive signal}

The final case to consider is a positive signal in GERDA whose rate cannot be described by neutrino physics only. As an example, we take a measured decay rate of $( T_{1/2}^{0\nu\beta\beta} )_{\rm C}=1.26\cdot 10^{27}$~years (well below the GERDA phase II limit), which would translate into an effective mass of $|m_{ee}|=0.045$~eV for an NME of 4.0. In case of a negative signal in KATRIN and Planck [and under the implicit assumption that there are no unknowns in cosmology which would affect $\Sigma$ in a way that it is not given by the expression in Eq.~\eqref{eq:covariance_observables} anymore], we would practically exclude inverted mass ordering, since the minimum value of $\Sigma$ in this case is $\Sigma_{\rm min}^{\rm IH}=0.1$~eV, which is just excluded by a negative signal from cosmology.

\begin{figure}[t]
\centering
\begin{tabular}{lr}
\includegraphics[width=0.43\textwidth]{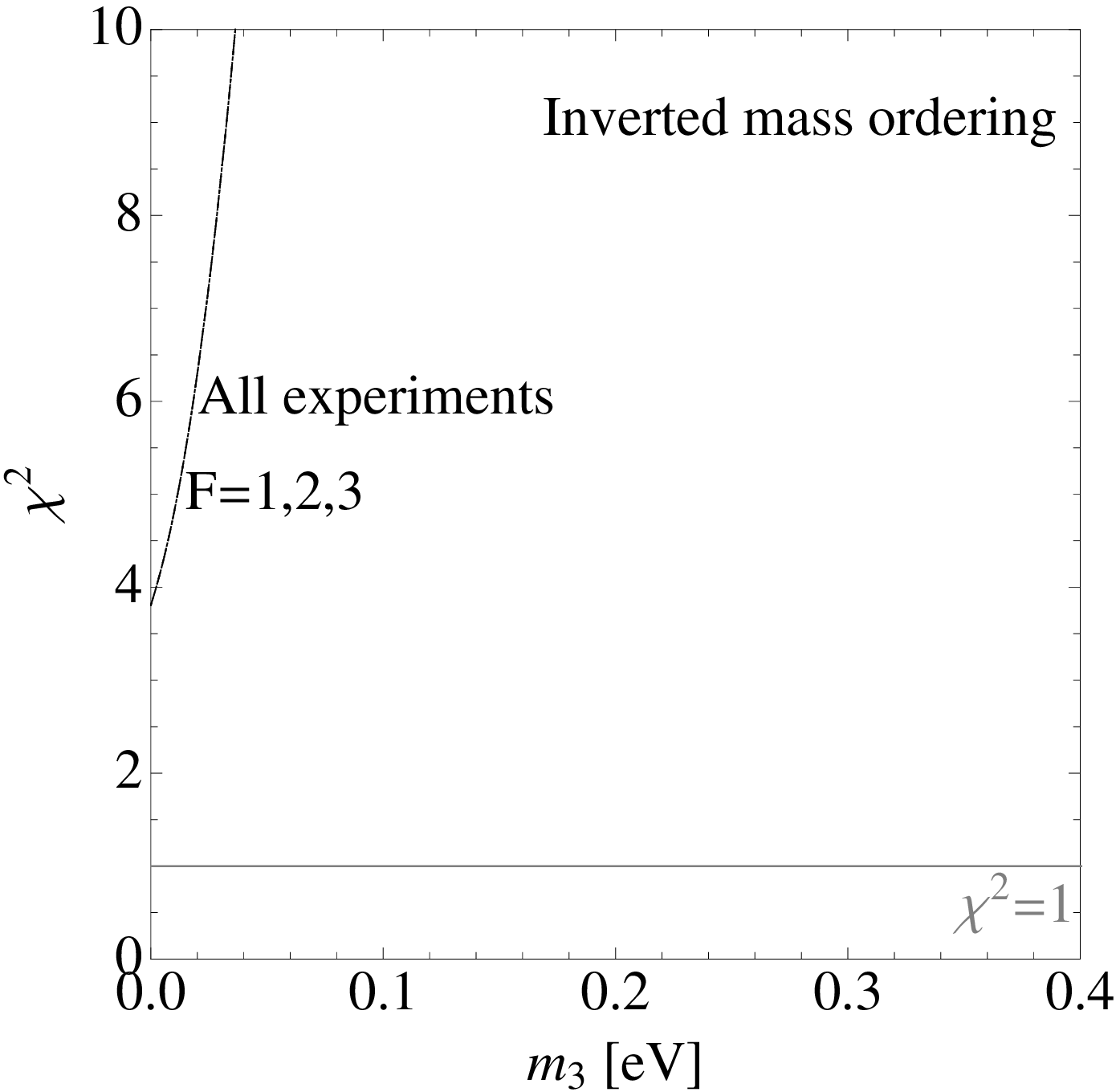}  &
\includegraphics[width=0.43\textwidth]{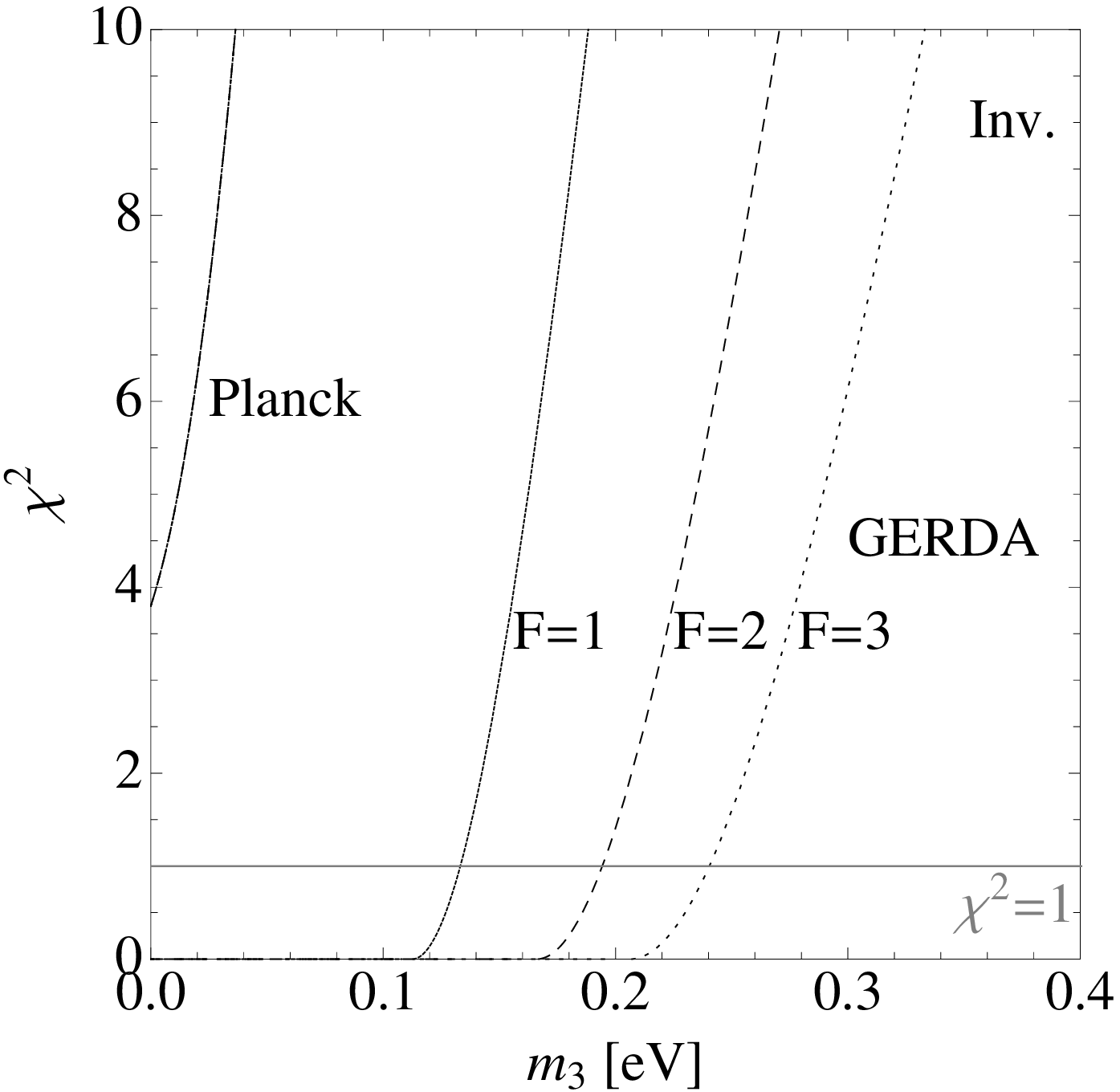}  \\
\includegraphics[width=0.43\textwidth]{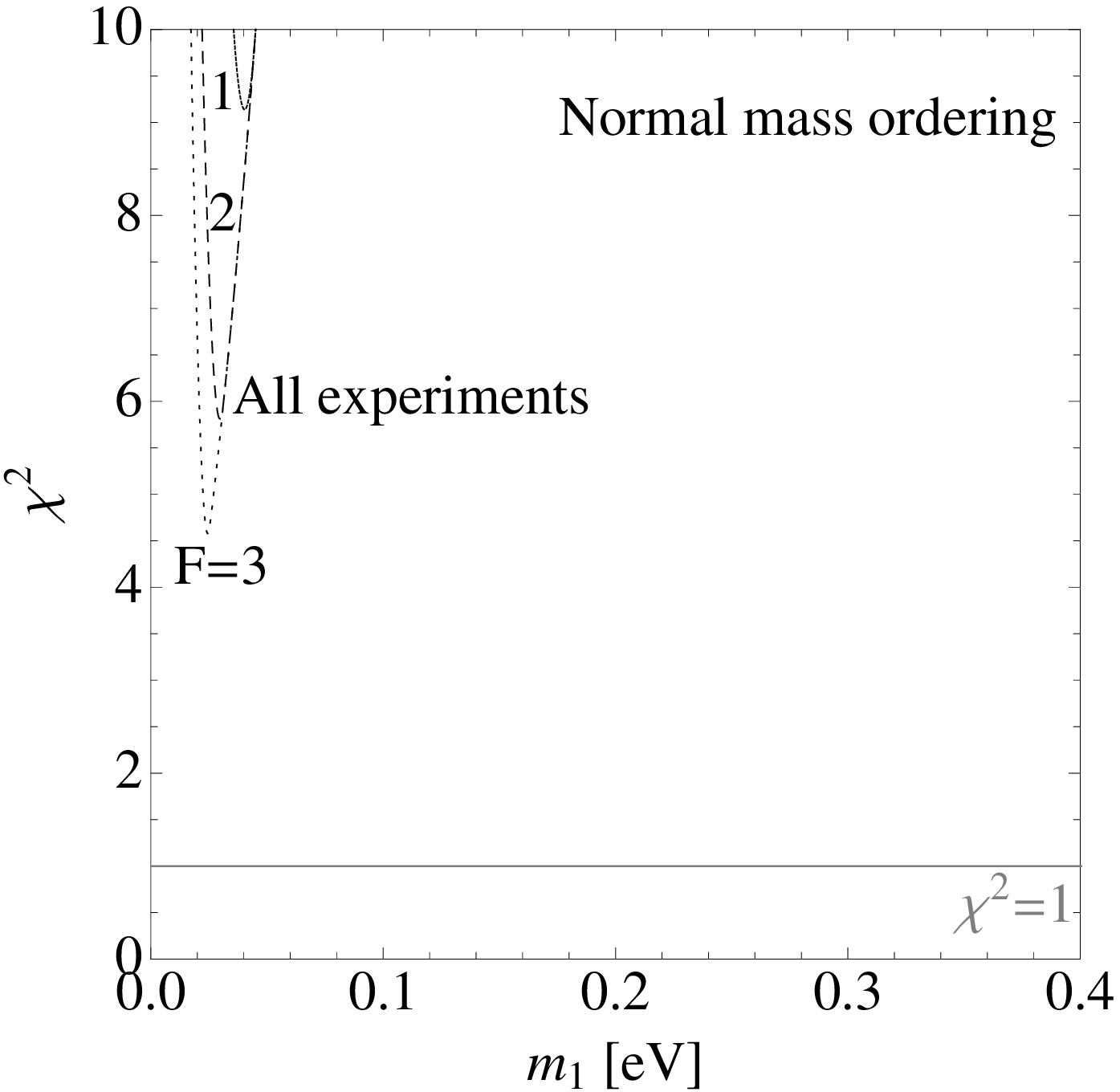} &
\includegraphics[width=0.43\textwidth]{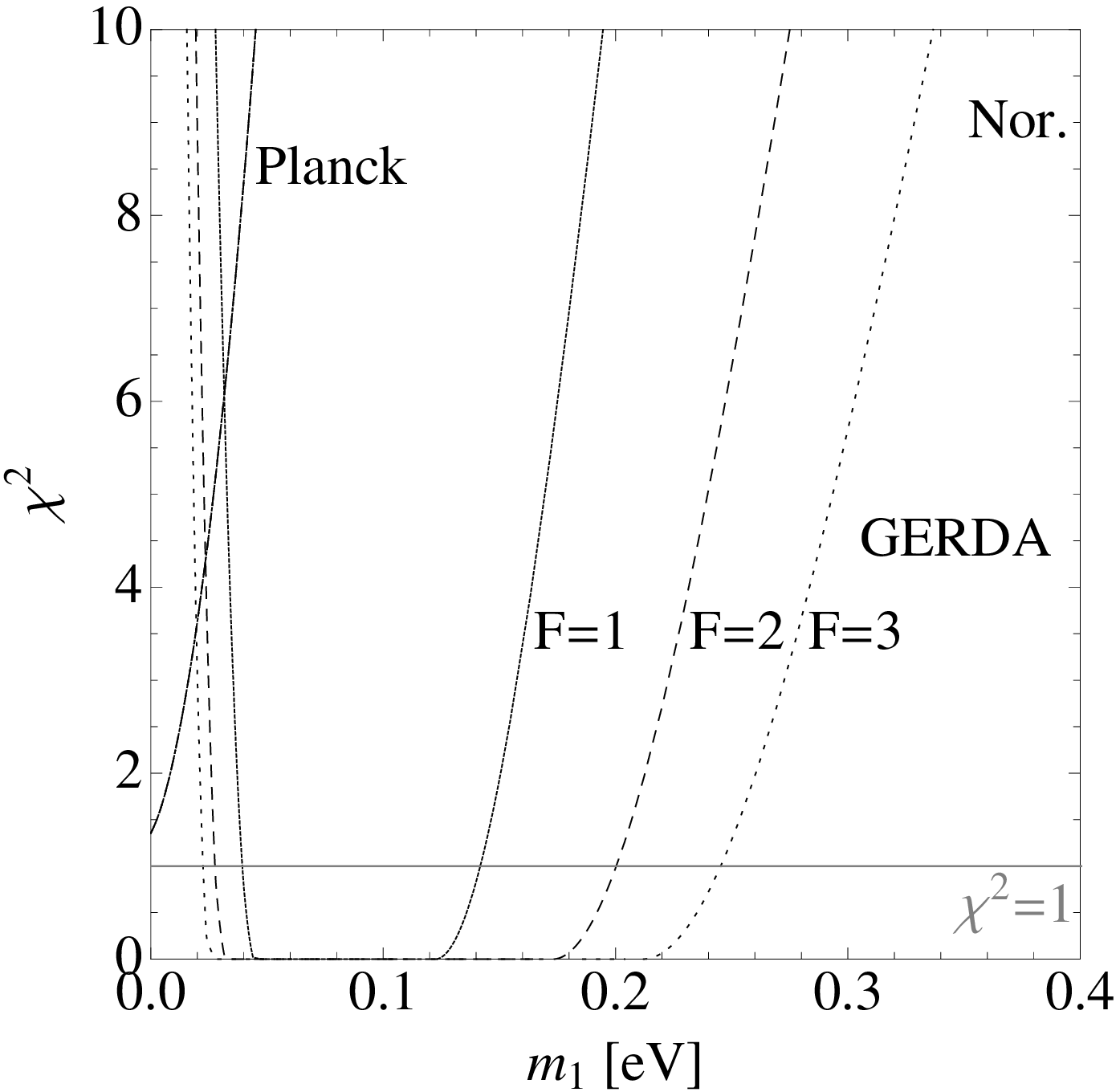}
\end{tabular}
\caption{\label{fig:chi2_C} The minimized $\chi^2$-function as a function of the smallest neutrino mass, which is $m_1$ for normal mass ordering and $m_3$ for inverted mass ordering.}
\end{figure}

In this case, the contribution of the light neutrino exchange to the decay rate would be smaller than the contribution from the effective operators by some factor. Then, under the assumption that the neutrino contribution can be completely neglected, one can actually derive a value (and, in principle,  allowed ranges), which serves as estimate for the coefficients $\eps_i$. Note that this would be in particular valid in case that the parameters are such that $|m_{ee}|$ is practically zero~\cite{Lindner:2005kr}. In Fig.~\ref{fig:chi2_C}, the minimized $\chi^2$-function as a function of the smallest neutrino mass is shown for normal and inverted mass ordering. We can see that, although we could in principle explain such a high \onbb\ rate for a certain range of the neutrino mass, this range does not fit with the additional information obtained by Planck which strongly disfavors the mass ranges required by GERDA. Accordingly, the minima of the full $\chi^2$-functions will be at least around 4, indicating that the results are inconsistent.

We will now use Eq.~\eqref{eq:T12_C} to estimate the magnitudes of the $\eps_i$'s. The \onbb\ decay rate $\Gamma_{\rm obs}$, cf.\ Eq.~\eqref{eq:GammaBas}, will translate into estimates of the sizes of the $\eps_i$'s,
\begin{equation}
|\eps_i |_{\rm obs} \simeq \frac{1}{\left| \mathcal{M}_i \right|} \sqrt{ \frac{\Gamma_{\rm obs}}{\tilde G_i\ \ln 2} }.
 \label{eq:eps_acc}
\end{equation}
Using the same formulas as in Sec.~\ref{sec:Analysis_A}, we obtain estimates in the case of scenario~C shown in Tab.~\ref{tab:results_C}.
\begin{table}[t]
\centering
\begin{tabular}{|c||c|c|c|c|c|c|}\hline
C & $|\eps_1|$ & $|\eps_2|$ & $|\eps_3^{LLz,RRz}|$ & $|\eps_3^{LRz,RLz}|$ & $|\eps_4|$ & $|\eps_5|$\\ \hline \hline
Estimate & $2.3\cdot 10^{-8}$ & $1.4\cdot 10^{-10}$ & $1.8\cdot 10^{-9}$ & $1.1\cdot 10^{-9}$ & $1.2\cdot 10^{-9}$ & $1.1\cdot 10^{-8}$\\ \hline
\end{tabular}
\caption{\label{tab:results_C} The resulting estimates for the magnitudes of the coefficients of the effective operators in scenario C, where $z=L$ or $R$.}
\end{table}

We conclude that one could, in principle, constrain the coefficients of the effective operators very well by simply assuming that the contributions from neutrino physics are completely subdominant. As to be expected, values slightly smaller than the bounds obtained in Ref.~\cite{Pas:2000vn} could be measured.

Accordingly, one can also derive the magnitude of the RPV-SUSY parameter in this scenario, which turns out from Tab.~\ref{tab:results_C} to be $\lambda'_{111}\simeq 0.074$. In principle, this method could be used to measure the amount of $R$-parity violation present.

\subsection{\label{sec:Caveat}On the importance of the knowledge of the nuclear physics}

There is an important caveat in our argumentation: In the above analyses, we have (in connection with the effective operators) always used the values for the NMEs derived from Refs.~\cite{Simkovic:2010ka,Fedor} (which update the ones given in Ref.~\cite{Pas:2000vn}). However, in reality these values will be uncertain. Different working groups might obtain different numbers, which could then be compared to the experimental results. A more advanced analysis that takes this into account has been performed in our scenarios~B and~C for the light neutrino exchange diagram. However, for the effective operators we do not want to guess some values of the NMEs that may be more or less likely, as long as no further calculations exist. One still has to keep in mind that exactly these computations might become very important in case that, after the next generation of experiments will have been completed, we would indeed end up in a situation that is similar to one of our scenarios.

The goal of this paper was to demonstrate how strong the bounds on point-like new physics can become in the case of a positive or negative future signal of \onbb. This has been achieved by the estimates obtained in Secs.~\ref{sec:Analysis_A}, \ref{sec:Analysis_B}, and~\ref{sec:Analysis_C}. However, this does not cure the problem of the relatively poor knowledge of the nuclear physics involved. Although this might not matter too much in a phenomenological analysis like the one presented here, it will be crucial for the true evaluation of future data -- a point that cannot be overemphasized. Hence, our results can very well serve as a guideline for how existing bounds on the effective operators will improve for different realistic situations. However, the numbers obtained suffer from the uncertainties involved in the computations of the nuclear matrix elements and can, accordingly, not be treated as strict statistical bounds.

\section{\label{sec:Conc}Conclusions}

We have investigated point-like operators contributing to neutrino-less double beta decay using the formalism of effective field theory. In order to focus on situations that can be tested in the near future, we have defined three scenarios, A, B, and C, which correspond to a negative double beta signal, a positive double beta signal that is consistent with complementary experimental results, and a positive double beta signal that is inconsistent with complementary experimental results, respectively. For each scenario, we have determined the bounds on (or estimates for) the strengths $\eps_i$ of each point-like operator that could contribute to the decay. Our scenario A is merely an update of bounds that have already been present in the literature, while our scenarios B and C have, to our knowledge, not been investigated before in this context. In case neutrino-less double beta decay is not detected at all, the current bounds on the $\eps_i$'s can be improved by roughly one order of magnitude. The most interesting case arises if a positive signal is consistent with all complementary results: Then, the possible room for further new physics contributions to double beta decay on top of the standard light neutrino exchange is essentially determined by the error in the measured rate, which is much smaller than the rate itself. In case of an inconsistent positive signal, one can even give estimates of the coefficients $\eps_i$. Our results do, however, suffer from the lack of knowledge of the underlying nuclear physics, which is a well-known problem in studies of neutrino-less double beta decay. Accordingly, the bounds that we have derived will be made much more robust once the nuclear matrix elements are known to a better precision. In that case, our bounds would depend directly on the experimental accuracy and could be significantly improved by future attempts to measure the rate of neutrino-less double beta decay more precisely.

\section*{\label{sec:Ack}\noindent Acknowledgments }

We would like to thank M.~D\"urr for useful discussions, and we are especially grateful to F.~\v{S}imkovic for kindly providing us with his values obtained for the partial NMEs. This work has been supported by the Swedish Research Council (Vetenskapsr{\aa}det), contract no.\ 621-2008-4210 (T.O.) and by the Royal Institute of Technology (KTH), project no.\ SII-56510 (A.M.).

\bibliographystyle{./apsrev}
\bibliography{./0nbb_EFT}

\end{document}